\documentclass{article}

\usepackage{arxiv}

\usepackage[utf8]{inputenc} 
\usepackage[T1]{fontenc}    
\usepackage{lmodern}        
\usepackage{hyperref}       
\usepackage{url}            
\usepackage{booktabs}       
\usepackage{amsfonts}       
\usepackage{nicefrac}       
\usepackage{microtype}      
\usepackage{graphicx}

\title{Estimating Disease-Free Life Expectancy based on Clinical Data
from the French Hospital Discharge Database}

\author{
  }

\providecommand{\tightlist}{%
  \setlength{\itemsep}{0pt}\setlength{\parskip}{0pt}}

\newlength{\cslhangindent}
\newlength{\csllabelwidth}
\newlength{\cslentryspacingunit} 
  {}%
  {\par}
\newenvironment{CSLReferences}[2] 
 {
  \setlength{\parindent}{0pt}
  \ifodd #1
  \let\oldpar\par
  \def\par{\hangindent=\cslhangindent\oldpar}
  \fi
  \setlength{\parskip}{#2\cslentryspacingunit}
 }%
 {}
\usepackage{calc}

\usepackage{booktabs}
\usepackage{longtable}
\usepackage[table]{xcolor}
\usepackage{array}
\usepackage{amsmath}
\usepackage{authblk}

\author[1,2]{Oleksandr Sorochynskyi}
\author[3]{Quentin Guibert}
\author[1,2]{Frédéric Planchet}
\author[4,5]{Michaël Schwarzinger}
\affil[1]{Laboratoire SAF EA2429, ISFA, University of Lyon, Université Claude Bernard Lyon 1, 69366 Lyon, France}
\affil[2]{Prim'Act Actuarial consulting firm, 42 avenue de la Grande Armée, 75017 Paris, France}
\affil[3]{CEREMADE, Université Paris-Dauphine, PSL University, CNRS 75016 Paris, France}
\affil[4]{Department of Prevention, Bordeaux University Hospital, 33000 Bordeaux, France}
\affil[5]{University of Bordeaux, INSERM, BPH, U1219, I-prev/PHARES, certified team under Ligue Contre le Cancer, CIC 1401, 33000 Bordeaux, France}
\usepackage{booktabs}
\usepackage{caption}
\usepackage{longtable}
\begin{document}
\maketitle

\begin{abstract}
The development of health indicators to measure healthy life expectancy
(HLE) is an active field of research aimed at summarizing the health of
a population. Although many health indicators have emerged in the
literature as critical metrics in public health assessments, the methods
and data to conduct this evaluation vary considerably in nature and
quality. Traditionally, health data collection relies on population
surveys. However, these studies, typically of limited size, encompass
only a small yet representative segment of the population. This
limitation can necessitate the separate estimation of incidence and
mortality rates, significantly restricting the available analysis
methods. In this article, we leverage an extract from the French
National Hospital Discharge database to define health indicators. Our
analysis focuses on the resulting Disease-Free Life Expectancy (Dis-FLE)
indicator, which provides insights based on the hospital trajectory of
each patient admitted to hospital in France during 2008-13. Through this
research, we illustrate the advantages and disadvantages of employing
large clinical datasets as the foundation for more robust health
indicators. We shed light on the opportunities that such data offer for
a more comprehensive understanding of the health status of a population.
In particular, we estimate age-dependent hazard rates associated with
sex, alcohol abuse, tobacco consumption, and obesity, as well as
geographic location. Simultaneously, we delve into the challenges and
limitations that arise when adopting such a data-driven approach.
\end{abstract}


\hypertarget{introduction}{%
\section{Introduction}\label{introduction}}

Over the last century, life expectancies have significantly increased.
However, this increase has also been accompanied by a rise in the
duration of life spent in a state of dependency (Fries 1980; Gruenberg
2005). This underscores the importance of health indicators, such as
Healthy Life Expectancy (HLE), in monitoring the overall health of a
population. HLE is an umbrella term for a family of health indicators
that calculate the expected number of years lived in various health
states.

HLE are utilized at all levels of policymaking, from international to
local. Organizations such as the World Health Organization (WHO) and the
European Union (EU) incorporate health indicators---Healthy Life
Expectancy (HALE) and Healthy Life Years (HLY), respectively--into their
frameworks for assessing population health (World Health Organization
2023; Bogaert et al. 2018). Another example is Japan, which has
prioritized health as a key policy objective in recent years (Abe 2013).

Despite the consensus on the importance of health indicators, no
universally used definition of health emerged (Chapter 1, Jagger et al.
(2020)). The complexity of defining a useful health concept and the
multiplicity of existing health concepts and methods to calculate them
are well-documented (Kim et al. 2022). However, one aspect appears to
remain invariant: the use of surveys.

Surveys are the main source of data on health status of a population
(Chapter 5, Jagger et al. (2020)). Unlike mortality data, that are
already available from national statistics agencies who collect it for
administrative purposes, health data are harder to come by, and surveys
provide the most readily available means of doing so. The use of survey
data necessarily imposes limits on the data collected. For one, the cost
of surveys limits the sample sizes. Constructing survey instruments to
be comparable over large areas is challenging (Robine 2003). Moreover,
self-evaluation of health which is influenced by various factors and can
therefore be biased (Kempen et al. 1996; Krause and Jay 1994; Peersman
et al. 2012). Finally, survey data also does not provide reliable
mortality data.

At the same time, the introduction of electronic health records (EHRs)
and international diagnostic harmonization has enabled the collection of
medical information across large populations, with datasets like the
United States' National Hospital Care Survey, the United Kingdom's
Hospital Episode Statistics database, and Denmark's National Patient
Registry. In this paper, we focus on a subset of the French National
Hospital Discharge database (Programme de Médicalisation des Systèmes
d'Information). These data cover all hospital discharges from 2010 to
2013 for adults aged 50 and older, and cover, after all exclusions, 10
million unique patients. Each discharge contains the main discharge
diagnosis, coded using ICD-10, a standardized international
classification of diagnoses (World Health Organization 2015), as well as
some demographic information on the patient.

We propose using such large clinical databases to construct health
indicators. This proposed approach has many advantages for assessing the
health status of a population. First, the use of standardized discharge
diagnosis codes, like the ICD-10, simplifies and reinforces
cross-regional and temporal comparisons. Second, involvement of
healthcare professionals in diagnosis minimizes biases associated with
self-assessment. As the entire population is included, the database can
provide a longitudinal view over a lifetime of diagnoses to create a
comprehensive health picture. Finally, the individual-level data that
contains information on both morbidity and mortality avoids the need for
aggregating and allows for more nuanced analysis, promising a more
profound view of health.

Nonetheless, the clinical view of health has inherent limitations.
First, a clinical view of health corresponds necessarily to a negative
concept of health, considered inadequate by some (Chapter 1, Jagger et
al. (2020)). In clinical settings, the focus is diagnosis and treatment,
not holistic health assessment. This divergence yields several notable
consequences (Euro-REVES et al. 2000). For instance, preventive measures
can avert certain conditions without the need for formal diagnosis.
Another concern associated with the clinical perspective is its reliance
on healthcare access levels (Sanders 1964). Moreover, the same diagnosis
can have varying effects on different individuals. A disease may or may
not lead to impairment or disability. For example, two people
experiencing a stroke may face different outcomes, a subtlety that may
not be accounted for by diagnoses alone. Finally, clinical data
represents only a part of the population. Therefore, producing estimates
representative of the general population is challenging and requires
additional assumptions to correct the selection bias causes by the
hospital admission. Even with these limitations we believe clinical data
can provide a complimentary view of population health.

In this paper, we develop a novel approach to constructing health
indicators from the family of Disease-Free Life Expectancy-type
indicators using clinical data. Most literature using Dis-FLE focuses on
a family of diseases: Lagström et al. (2020) focuses on cardiometabolic
disease, while Head et al. (2019) and Stenholm et al. (2017) focus
chronic conditions such as cardiovascular disease, stroke, cancer,
respiratory disease, and diabetes. We aim to broaden the considered
diseases even further by simultaneously considering a wide range of
diseases that can lead to severe health deterioration or mortality. This
approach helps mitigate bias associated with tracking a single specific
condition to assess changes in health status. The obtained Dis-FLE
indicator is then compared to HLY.

A second contribution of this paper is to utilize information from
clinical data to assess variations in Dis-FLE based on different risk
factors. In contrast with the traditional Sullivan's method, our
approach based on individual data and a Cox model is able to assess the
effect of different covariates. To do so, we consider the age-dependent
impact of sex, behavioral risk factors, and the interactions thereof. We
also take into account the region of residence. In doing so, we can not
only present an estimate of Dis-FLE for each stratum but to gauge to
some extent its main determinants.

The rest of this paper, is structured as follows. Section \ref{sec:data}
introduces the data used. Section \ref{sec:stat_methods} describes the
statistical methods used to construct health indicators. The results are
presented in Section \ref{sec:results}, which is broken into three
subsections. Section \ref{sec:hly} presents the Dis-FLE estimations, and
compares them to HLY. Section \ref{sec:cox} analyzes Dis-FLE
determinants using a Cox model. Section \ref{sec:discuss} concludes by
providing a discussion of the approach and the results.

\hypertarget{data}{%
\section{\texorpdfstring{Data \label{sec:data}}{Data }}\label{data}}

\hypertarget{description}{%
\subsection{Description}\label{description}}

This study uses a subset of the French National Hospital Discharge
database, PMSI, that covers 2010 to 2013. These data cover all hospital
visits in Metropolitan France during the observation period. Only
hospital stays of people ages 50 and up are included in this subset. For
this age category, over 75\% of the general population appear in the
database. These data were previously used in Schwarzinger et al. (2018)
and Schwarzinger (2018). The first references provides the ICD-10
(International Classification of Diseases, 10th Revision) codes which
were used to identify conditions as well as some risk factors. The
second reference is however in French, we therefore include a brief
description here.

For each patient, the data include a series of discharge dates and the
associated diagnosis. These enable us to track individual health
trajectories over time. A severe condition in this study should be
understood as a medical syndrome encompassing multiple diseases or
evolving stages with a high risk of disability or death. A typical
example of a severe condition is `dementia,' which includes Alzheimer's
disease and related conditions, i.e., all causes of cognitive loss of
autonomy (Schwarzinger 2018). The notion of disease-free used in this
paper is based on these severe conditions.

Some exclusion criteria are applied to construct health indicators
relative to a healthy population, in terms of the selected conditions.
These criteria are adapted from Schwarzinger (2018). Firstly, we exclude
patients observed for any of the severe conditions used to define the
healthy population during the period 2008-2010. Here, we assume that
individuals within the general population that did not appear in a
hospital during this 2-year period for any of the selected conditions,
whether for an initial consultation or follow-up for a chronic disease,
are in good health, i.e., they are not affected by the consequences of
these conditions. Thus, this procedure allows us to obtain, as of
January 1, 2010, a selected population without any history of severe
conditions for over 2 years. Additionally, we exclude 914,595
individuals hospitalized from 2008 to 2013 for certain chronic
conditions (e.g., birth defects, HIV infection, psychiatric disorders,
etc.) In this regard, we observe that 375,579 (41\%) of these
individuals are already included in the first exclusion group. After
exclusions, data include almost 30 million hospital visits and over 10
million unique individuals over the observation period, see Table
\ref{tab:filters-schwarz} for the details of the exclusion criteria.

Table \ref{tab:individu-cols} describes the information available for
individual patients. Basic demographic information is available : year
of birth, sex and approximate place of residence, i.e., the departement
of residence among the 96 official French administrative departments
over the period 2008-2013. To enable the estimation of regular survival
functions, a fictitious birthdate is imputed for each patient. Three
lifestyle risk factors are inferred from hospital data and prior
diagnoses : active tobacco smoking, alcohol use disorders and obesity
(body-mass index \(\geq\) 30 kg/m\(^2\)). Each risk factor is classified
into 3 categories : 0, 1 and 2; 0 being the absence of risky behavior
(Schwarzinger et al. 2018). It should be noted that alcohol or tobacco
consumption is defined based on medical codes rather than on patients'
self-reporting. Therefore, these variables capture a relatively severe
exposure to these factors. Information on education levels and
immigration status is a commune-level proxy (i.e., it represents the
education/immigration levels of the commune of residence not of the
individual) based on INSEE data. Individual-level information is
collected on the first hospital visit, and is assumed to be constant
over time.

\begin{table}

\caption{\label{tab:individu-cols}Description of individual patient data.}
\centering
\begin{tabular}[t]{lll>{\raggedright\arraybackslash}p{16em}}
\toprule
Column & Precision & Possible values & Description\\
\midrule
ID & individual & positive integers & Anonymized identifier\\
Alcohol & individual & 0, 1, 2 & Alcohol use disorder, grouped into three classes in increasing order: '0' for the absence of Alcohol use disorder, '1' for mental and behavioural disorders due to former or current chronic harmful use of alcohol (ICD-10: F10.1–F10.9, Z50.2) including alcohol abstinence (ICD-10: F10.20–F10.23), '2' chronic diseases attributable to alcohol use disorders (e.g., WernickeKorsakoff syndrome, end-stage liver disease and other forms of liver cirrhosis, epilepsy, and head injury)\\
Obesity & individual & 0, 1, 2 & Obesity, grouped into three classes in increasing order: '0' boby mass < 30 kg/m$^2$, '1' boby mass $\geq$ 30 kg/m$^2$ and
  < 40 kg/m$^2$, '2' boby mass $>$ 40 kg/m$^2$.\\
Smoker & individual & 0, 1, 2 & Smoking, grouped into three classes in increasing order : '0': no disorder due to tobacco use recorded, '1': mental and behavioral disorders due to tobacco use (ICD-10: F17), '2':
  mental and behavioral disorders due to tobacco use (ICD-10: F17) and Chronic Obstructive Pulmonary Disease (ICD-10: J44.9).\\
Department & individual & `01' to `96' & Department of residence (Metropolitan France)\\
\addlinespace
Immigration & postal code & 0, 1, 2, 3 & Proportion of foreign nationals, grouped into quartiles, proxy for immigration status\\
Education & postal-code & 0, 1, 2, 3 & Proportion population with higher education, grouped into quartiles, proxy for education\\
Sex & individual & `M' or `F' & M:male, F:female\\
Year of birth & individual & integer & Year of birth\\
\bottomrule
\end{tabular}
\end{table}

We define disease-free as the absence of new events described in Table
\ref{tab:pathol}. This choice is motivated by previous research on this
dataset (Guibert, Planchet, and Schwarzinger 2018a; Schwarzinger 2018).
These previous works define disability much more narrowly, considering
only two events, ``Physical dependence'' and ``Severe dementia''.
Physical dependence is defined as bedridden state. Schwarzinger (2018)
established that this definition of disability aligns with severe
disability as measured by activities of daily living (ADLs). In our
study, we aim to broaden our definition of disability to encompass all
identified severe events, bringing it closer to a less severe level of
activity limitation, similar to the concept of Global Activity
Limitation Instrument (GALI), the measure of disability used for HLY.

The approach used to define disease-free in this study is distinctive,
as it covers essentially all diseases that increase the risk of death.
Indeed, this list almost exhaustively covers the various causes of death
with 98\% of the 1,774,703 deaths in the hospital from 2008 to 2013
(Schwarzinger 2018). Moreover, we believe that including such a wide
range of diseases brings the resulting Dis-FLE closer to a
general-notion of population health.

It is worth noting that the list of events used to define disability
employed in this study was not explicitly designed to mirror existing
health measures, such as GALI. Instead, it represents the closest
available approximation using this data, based on our knowledge. While
this approach allows us to assess the merits of using clinical data, it
is important to recognize that the indicator used may not capture the
same aspects of health as existing health indicators.

\begin{table}

\caption{\label{tab:pathol}List of 36 severe conditions requiring hospital care and considered incompatible with good health and number of times the event was observed during the 2010-2013 period.}
\centering
\begin{tabular}[t]{>{\raggedright\arraybackslash}p{20em}r}
\toprule
Event description & Number of events observed\\
\midrule
Heart failure (including cardiac arrest) & 967 187\\
Rhythm disorder: 1 Atrial fibrillation & 705 528\\
Peripheral arterial disease (aorta, digestive system, kidney, amputation) & 531 657\\
Anemia: 1 Blood transfusion & 502 472\\
Chronic kidney disease & 335 038\\
Digestive complication: 1 Hemorrhage (any cause) & 333 720\\
Septicemia (any cause) & 270 932\\
Thromboembolic disease & 265 323\\
Acute respiratory failure & 235 699\\
Digestive complication: 1 Obstruction (any cause) & 231 039\\
Stroke: 1 Ischemic (less severe) & 222 098\\
Acute kidney failure & 220 118\\
Breast cancer & 205 026\\
Metabolic disease (other than diabetes, dyslipidemia) & 201 431\\
Lung cancer & 194 172\\
Chronic respiratory failure (including respiratory arrest) & 185 282\\
Prostate cancer & 184 291\\
Severe dementia & 178 670\\
Cancer with poor prognosis & 161 046\\
Ischemic Heart Disease: 1 Heart attack (stent, surgery) & 160 190\\
Trauma: 1 Skull & 159 856\\
Colorectal cancer & 151 427\\
Epilepsy (and other convulsions) & 131 300\\
Hemopathy (lymphoma) & 130 476\\
Parkinson's disease (and other extrapyramidal syndromes) & 128 533\\
Endocrine disease (other than thyroid) & 111 232\\
Digestive complication: 1 Peritonitis (any cause) & 103 674\\
Cancer with good prognosis & 99 185\\
Digestive complication: 1 Stoma (any cause) & 89 719\\
Cirrhosis: 1 Decompensated & 89 701\\
Physical dependence (bedridden state without dementia) & 87 736\\
Stroke: 1 Hemorrhagic (more severe) & 79 028\\
ORL Esophageal cancer & 72 482\\
Trauma: 2 Severe (non-skull) & 69 537\\
Other neurological disease & 57 739\\
Rare diseases at risk of dementia (multiple sclerosis, normal-pressure hydrocephalus, encephalitis) & 45 083\\
Death from any cause & 569 941\\
\bottomrule
\end{tabular}
\end{table}

\hypertarget{summary-statistics}{%
\subsection{\texorpdfstring{Summary statistics
\label{sec:stats}}{Summary statistics }}\label{summary-statistics}}

Table \ref{tab:demo_stats} gives summary statistics of the population
under study. Women represent a larger proportion of the population, for
two reasons. First, women tend to live longer and second, a higher
proportion of women has visited hospitals.

The exact age in years is used as the timescale for the analysis. The
exact age is the number of years since birth, including the fractional
part. Individuals are considered exposed from their 50th birthdays to
the first adverse event, within the period from 2010 to 2013, the
observation period.

For all three risk behaviors over 85\% of the population are in category
0, i.e., absence of any risk factor. This reflects the fact that risk
factors represent relatively severe cases of each behavior. The
immigration and education variables are grouped into quartiles.

\begin{table}\begin{longtable}{l|rrr}
\caption{
\label{tab:demo_stats}Descriptive statistics of information
        available for the analysis.
} \\ 
\toprule
\multicolumn{1}{l}{} & \multicolumn{3}{c}{Sex} \\ 
\cmidrule(lr){2-4}
\multicolumn{1}{l}{} & Female & Male & Entire population \\ 
\midrule
\multicolumn{4}{l}{Number of indiduals} \\ 
\midrule
\hspace*{25px} n & 5 849 485 & 4 761 144 & 10 610 629 \\ 
\midrule
\multicolumn{4}{l}{Age at start of exposure} \\ 
\midrule
\hspace*{25px} Median (IQR) & 64.9 (56.3—76.2) & 62.2 (55.3—72.1) & 63.5 (55.8—74.4) \\ 
\midrule
\multicolumn{4}{l}{Exposure (years)} \\ 
\midrule
\hspace*{25px} Median (IQR) & 2.1 (1.1—3.1) & 2.1 (1.0—3.1) & 2.1 (1.0—3.1) \\ 
\midrule
\multicolumn{4}{l}{Obesity} \\ 
\midrule
\hspace*{25px} Category  0 (\% of pop.) & 5 333 571 (91.2\%) & 4 375 255 (91.9\%) & 9 708 826 (91.5\%) \\ 
\hspace*{25px} Category  1 (\% of pop.) & 420 360 (7.2\%) & 338 803 (7.1\%) & 759 163 (7.2\%) \\ 
\hspace*{25px} Category  2 (\% of pop.) & 95 554 (1.6\%) & 47 086 (1.0\%) & 142 640 (1.3\%) \\ 
\midrule
\multicolumn{4}{l}{Alcohol} \\ 
\midrule
\hspace*{25px} Category  0 (\% of pop.) & 5 762 344 (98.5\%) & 4 520 484 (94.9\%) & 10 282 828 (96.9\%) \\ 
\hspace*{25px} Category  1 (\% of pop.) & 17 370 (0.3\%) & 38 044 (0.8\%) & 55 414 (0.5\%) \\ 
\hspace*{25px} Category  2 (\% of pop.) & 69 771 (1.2\%) & 202 616 (4.3\%) & 272 387 (2.6\%) \\ 
\midrule
\multicolumn{4}{l}{Smoking} \\ 
\midrule
\hspace*{25px} Category  0 (\% of pop.) & 5 559 858 (95.0\%) & 4 224 623 (88.7\%) & 9 784 481 (92.2\%) \\ 
\hspace*{25px} Category  1 (\% of pop.) & 9 817 (0.2\%) & 29 173 (0.6\%) & 38 990 (0.4\%) \\ 
\hspace*{25px} Category  2 (\% of pop.) & 279 810 (4.8\%) & 507 348 (10.7\%) & 787 158 (7.4\%) \\ 
\midrule
\multicolumn{4}{l}{Immigration} \\ 
\midrule
\hspace*{25px} Quartile  0 (\% of pop.) & 888 302 (15.2\%) & 738 362 (15.5\%) & 1 626 664 (15.3\%) \\ 
\hspace*{25px} Quartile  1 (\% of pop.) & 1 234 883 (21.1\%) & 945 294 (19.9\%) & 2 180 177 (20.5\%) \\ 
\hspace*{25px} Quartile  2 (\% of pop.) & 1 613 637 (27.6\%) & 1 249 331 (26.2\%) & 2 862 968 (27.0\%) \\ 
\hspace*{25px} Quartile  3 (\% of pop.) & 2 112 663 (36.1\%) & 1 828 157 (38.4\%) & 3 940 820 (37.1\%) \\ 
\midrule
\multicolumn{4}{l}{Education} \\ 
\midrule
\hspace*{25px} Quartile  0 (\% of pop.) & 1 397 459 (23.9\%) & 1 126 502 (23.7\%) & 2 523 961 (23.8\%) \\ 
\hspace*{25px} Quartile  1 (\% of pop.) & 1 563 617 (26.7\%) & 1 243 339 (26.1\%) & 2 806 956 (26.5\%) \\ 
\hspace*{25px} Quartile  2 (\% of pop.) & 1 472 192 (25.2\%) & 1 182 230 (24.8\%) & 2 654 422 (25.0\%) \\ 
\hspace*{25px} Quartile  3 (\% of pop.) & 1 416 217 (24.2\%) & 1 209 073 (25.4\%) & 2 625 290 (24.7\%) \\ 
\bottomrule
\end{longtable}
\end{table}

Table \ref{tab:fdr_cor} shows correlations between presence of risk
factors. Correlations for risk factors are calculated on the indicator
variables for any category risk factor, i.e., category 1 and 2 risk
factors are grouped together. For Education and Immigration, the numeric
0-based quartile is taken. All correlations are highly significant (p
\textless{} 0.001), but most are small. There is a correlation between
alcohol consumption and smoking. The correlation between immigration and
education is hard to interpret as it is likely a reflection of postal
codes rather than individuals.

\begin{table}\begin{longtable}{l|rrrr}
\caption{
\label{tab:fdr_cor}Correlations between risk factors.
        Only the presence of each risk factor is considered, ignoring
        categories.
} \\ 
\toprule
\multicolumn{1}{l}{} & Education & Immigration & Obesity & Smoking \\ 
\midrule
Alcohol & 0.02 & 0.01 & 0.03 & 0.22 \\ 
Education &  & 0.24 & 0.04 & 0.03 \\ 
Immigration &  &  & 0.00 & 0.01 \\ 
Obesity &  &  &  & 0.09 \\ 
\bottomrule
\end{longtable}
\end{table}

\hypertarget{methods}{%
\section{\texorpdfstring{Methods
\label{sec:stat_methods}}{Methods }}\label{methods}}

\hypertarget{statistical-tools}{%
\subsection{Statistical tools}\label{statistical-tools}}

In our study, we employ two types of models: the Kaplan-Meier estimator
for survival curves and the Cox proportional hazards model. See, for
example, Klein et al. (2016) for general background on survival models.
The Kaplan-Meier estimator stratifies the population and calculates
survival curves separately for each stratum. In contrast, the Cox model
takes into account all available data and covariates simultaneously.
Furthermore, the Cox model offers a method for estimating a survival
curve based on the covariates in question. Both methods rely on the
assumption that the censoring time is independent of both the exit time
and the covariables.

The Kaplan-Meier survival function estimator at time \(t\) is given by :
\begin{equation}
\hat{S}(t) = \prod_{\{i:t_i \leq t\}} \left(1 - \frac{d_i}{n_i}\right),
\end{equation} where:

\begin{itemize}
  \item $t_i$ is the observed event time for the $i^\text{th}$ observation,
  \item $d_i$ is the number of non-censored events at $t_i$,
  \item $n_i$ is the number of individuals at risk just before $t_i$.
\end{itemize}

To obtain stratum-specific survival curves the estimator is calculated
independently for each subset of data.

The Cox model, in contrast to Kaplan-Meier, is a regression model as it
attempts to establish a link between covariables and the survival time.
It does so by assuming that all observations share the same baseline
hazard function, \(\lambda_0(t)\), that is scaled by the covariables.
The Cox model estimates the hazard function as : \begin{equation}
\hat{\lambda}(t | \mathbf{X}) = \lambda_0(t) e^{\mathbf{X}\beta},
\end{equation} where \(\mathbf{X}\) is the design matrix and \(\beta\)
are Cox model coefficients. To obtain survival curve estimates, we also
need to estimate the baseline hazard function \(\lambda_0\) or
equivalently its cumulative counterpart
\(\Lambda_{0}(t) = \int_0^t \lambda_0(u)\,du\). We use the Breslow
estimate for the cumulative baseline hazard function : \begin{equation}
\hat{\Lambda}_{0}(t) = 
  \sum_{\{i : t_i \leq t\}}
      \frac{\delta_i}{
        \sum_{k \in \mathcal{R}_i} e^{\mathbf{X}_k\hat{\beta}}}.
\end{equation} Here, \(\delta_i\) represents the event indicator (1 if
the event occurred, 0 if censored). The summation is performed over all
events \(i\) where exit time \(t_i \leq t\). The denominator calculates
the risk set contribution for observations still under risk at \(t_i\),
with \(\mathcal{R}_i = \{j : t_j \geq t_i\}\) and \(\hat{\beta}\) the
maximum likelihood estimator of Cox model coefficients. Overall, for the
Cox model, the survival function is estimated using \begin{equation}
\hat{S}(t | \mathbf{x}) = \exp\left(-\int_0^t e^{\mathbf{x}\hat{\beta}} \, d\hat{\Lambda}_0(u)\right).
\end{equation} This basic variant of the Cox model assumes that the
conditional hazard functions are all proportional to a base hazard
function, This assumption is not satisfied for these data. For this
reason we use a variant of the model that allows the hazard ratio to
vary over time, in this case age, thus reducing non proportionality
\(\lambda(t, \mathbf{X}) = \lambda_{0}(t)e^{\mathbf{X}\beta(t)}\)
(Martinussen and Scheike 2006). This procedure requires duplicating each
observation for every change in \(\beta(t)\). For this reason, instead
of using every event time we choose a coarse grid of ages : steps of 2
years from 50 to 100. This results in step-function estimate for
coefficients with time dependent effect. We use a natural spline basis
to estimate \(\beta(t)\). In the rest of the paper we refer to these
time-dependent coefficients as age-dependent as age is the timescale
used for this model.

Initial data wrangling is done in SAS. Further data treatment and
analysis is done in R (R Core Team 2022). The Kaplan-Meier survival
curves and Cox model was estimated using methods from the
\texttt{survival} package (Therneau 2023). The procedure
\texttt{survSplit} from the \texttt{survival} package is used to split
observations over time, as required to estimate age-dependent effects.
The splines are implemented using \texttt{nsk} function from the same
package.

\hypertarget{statistical-modeling}{%
\subsection{Statistical modeling}\label{statistical-modeling}}

We analyze health as a censored life duration without disease. Our
estimation approach relies on the use of survival models. The observed
individual disease-free life durations, denoted \(T\), are subject to
right censoring and left truncation linked to the observation period.
The truncation and censoring dates are assumed to be independent of
\(T\). An important assumption that we make is that the conditions
selected to define Dis-FLE are supposed to be severe enough to require
hospital care. Thus, we consider that the information loss related to
patients with these conditions but not observed in hospital induces
limited bias.

The duration studied is the disease-free survival which we define as the
time between the start of the observation (either 2010/01/01 or the 50th
birthday, whichever comes later) end the end of observation (either
2013/12/31 or date of death or censoring, whichever comes first).
Censoring can be due to the end of the observation period on 2013/12/31
or due to being lost to follow-up. For Kaplan-Meier only sex is used to
stratify the population, whereas for the Cox model uses many variables
as covariables are used, as described in the end of this section. Both
methods allow estimating survival curves.

We view Dis-FLE as the expected value of the disease-free survival
distribution conditional on attaining a certain age. The disease-free
survival distribution can either be estimated using either Kaplan-Meier
or Cox model. Formally, if \(S\) is the estimate of the survival curve
of \(T\), then the restricted conditional expectation is
\(\text{Dis-FLE}(t) = \mathbb{E}(T-t | T > t)\), for \(t \geq 50\), and
can be calculated by \[
\int_{t}^{t_{\text{max}}} \frac{S(u)}{S(t)}\, du,
\] where \(t_{\text{max}}\) is the maximum assumed age. We set
\(t_{\text{max}}\) to 100, the largest age in the INSEE age pyramid used
in whole population adjustment (see Section \ref{sec:whole_pop_adj}).
Setting a maximal age is one way of dealing with the fact that survival
function does not reach 0 if the longest observation is censored.

However, given that both survival function estimators are step
functions, this formula reduces to a weighted sum. The formula used to
calculate Dis-FLE is given by : \[
\text{Dis-FLE}(t) =
  \sum_{i : t_{(i)} \geq t} \frac{\hat{S}(t_{(i)})}{\hat{S}(t)}(t_{(i+1)} - t_{(i)}),
\] where \(t_{(i)}\) are the unique, ordered, non-censored exit times
observed in the data, such that
\(t_{(1)} < t_{(2)} < \cdots < t_{(n)}\). We assume that this grid is
sufficiently small so that we have \(t_{(i)} = t\) for the first
\(i : t_{(i)} \geq t\). The first value of the Dis-FLE curve,
\(\text{Dis-FLE}(50)\), is the expected disease-free life duration at
50.

We first calculate and present sex-specific survival curves estimated
via Kaplan-Meier. We then calculate the corresponding
\(\text{Dis-FLE}(t)\) for all ages \(t \geq 50\). The main part of the
analysis is done using a Cox proportional hazards model.

The covariates used in the Cox model include sex, behavioral risk
factors, and geographical information. All terms of the Cox model are
described in Table \ref{tab:cox_terms}. Sex and all risk factors have
age-dependent coefficients. Age-dependent coefficients are obtained by
including in the model an interaction term between a natural spline as a
function of age and the age-dependent effect. The main effects (i.e.,
without interaction with age-dependent spline) are not included because
they would be colinear with the interaction effect. The relationship
with age is modeled using cubic natural splines with 8 degrees of
freedom, and with knots at the edges of observed values to prevent
linear extrapolation at the extremes. The interaction terms are modeled
as a constant offset of the main age-dependent effect.

\begin{table}\begin{longtable}{lll}
\caption{
\label{tab:cox_terms}Terms used in the Cox model.
} \\ 
\toprule
Term & Dependence on age & Reference value \\ 
\midrule
\multicolumn{3}{l}{Main effect} \\ 
\midrule
Obesity & Natural spline & Category 0 \\ 
Alcohol & Natural spline & Category 0 \\ 
Tobacco & Natural spline & Category 0 \\ 
Sex & Natural spline & Female \\ 
Department of residence & Constant & 78 - Yvelines \\ 
Immigration level & Constant & 1st quantile (lowest) \\ 
Education level & Constant & 1st quantile (lowest) \\ 
\midrule
\multicolumn{3}{l}{Interaction} \\ 
\midrule
Obesity * Alcohol & Constant & Both category 0 \\ 
Obesity * Tobacco & Constant & Both category 0 \\ 
Alcohol * Tobacco & Constant & Both category 0 \\ 
Sex * Obesety & Constant & Female, category 0 \\ 
Sex * Alcohol & Constant & Female, category 0 \\ 
Sex * Tobacco & Constant & Female, category 0 \\ 
\bottomrule
\end{longtable}
\end{table}

We usually avoid discussions of p-values, or significance tests, for two
reasons. First is practical, with such large data almost all comparisons
detect significant differences. Second is conceptional, data analyzed
exhaustively covers the studied population, therefore estimates are not
subject to sampling error.

40\% of observed individuals are randomly reserved for model validation,
which is shown in the Appendix. Indeed, the volume of data is more than
sufficient to estimate the model described above, as can be seen from
small standard errors of estimated coefficients.

\hypertarget{whole-population-adjustment}{%
\subsection{\texorpdfstring{Whole population adjustment
\label{sec:whole_pop_adj}}{Whole population adjustment }}\label{whole-population-adjustment}}

The Metropolitan French PMSI dataset analyzed in this article is limited
to individuals who have been hospitalized at some point, forming a
non-random sample of the broader French population. Consequently, any
calculations Dis-FLE within this sub-population yield a biased estimate
of the true general population indicator, rendering direct comparisons
impractical. To make a meaningful comparison with HLY, we make the
assumption that individuals not observed in PMSI are in good health and
adjust the exposure accordingly.

This disparity is not surprising given the substantial number of
individuals who have never been hospitalized. In 2010, France had 22.5
million individuals aged 50 and over (INSEE 2022), but only 10.5 million
observed in hospital and included in this study after various
exclusions. Indeed, hospitalization introduces selection bias that needs
to be corrected. There are two distinct and opposite sources of bias :

\begin{enumerate}
\def\labelenumi{\arabic{enumi}.}
\tightlist
\item
  the population included in the PMSI is, on average, in worse health
  than the general population since they required hospitalization and
\item
  exclusions applied to the original PMSI data should result in a study
  population that is healthier than the PMSI population.
\end{enumerate}

Of these two effects the first one is stronger, and is the one we
attempt to correct using this adjustment.

Let \(l_{x,k}\) represent the population aged \(x\) on January the 1st
of year \(k\). The adjustment is made by introducing
\(l_{2010-c,2010}^\text{INSEE} - l_{2010-c,2010}^\text{PMSI}\)
artificial data points without any disease, corresponding to individuals
not observed in the PMSI on 2010/01/01, for each observed cohort \(c\)
(year of birth) and separately for each sex, notation notwithstanding.
These individuals are then censored at the end of years 2010 through
2013 as needed to align the exposure with INSEE data.

It's important to note that the assumption on which this adjustment is
made---that individuals not present in the PMSI database are alive and
in good health---is not universally satisfied : (1) it disregards the
subpopulation initially included in PMSI but later excluded for this
study, and (2) it does not account for rare events missed by PMSI. The
first point is handled by scaling the observed population size,
\(l_{2010-c,2010}^\text{PMSI}\), to the pre-exclusion levels before
calculating by how much the exposure needs be increased to match the
entire French population. This done to avoid re-adding the excluded
population back in as healthy observations. The scaling factor
corresponds to a 40\% increase and is simply the ratio between the
population before exclusions and after :
\(18\,440\,022 / 13\,170\,355 \approx 1.40\), both values come from
Table \ref{tab:filters-schwarz}. The use of the scaling factor is a
simplification as it assumes that the exclusions had proportionally the
same effect on all ages. The search for an adjustment to correct the
selection bias caused by the use of clinical data is a delicate topic
that is outside the scope of this paper. The second point cannot be
handled easily.

It's essential to emphasize that this adjustment can only be applied
when considering sex as the sole covariate. We cannot employ this
adjustment for the Cox model since we lack individual-level information
on covariates for the entire population. Therefore, Cox model should be
interpreted as estimating the risk relative to the hospitalized
population.

\hypertarget{results}{%
\section{\texorpdfstring{Results
\label{sec:results}}{Results }}\label{results}}

\hypertarget{dis-fle-and-comparison-to-eurostats-hly}{%
\subsection{\texorpdfstring{Dis-FLE and comparison to Eurostat's HLY
\label{sec:hly}}{Dis-FLE and comparison to Eurostat's HLY }}\label{dis-fle-and-comparison-to-eurostats-hly}}

\begin{figure}
\centering
\includegraphics{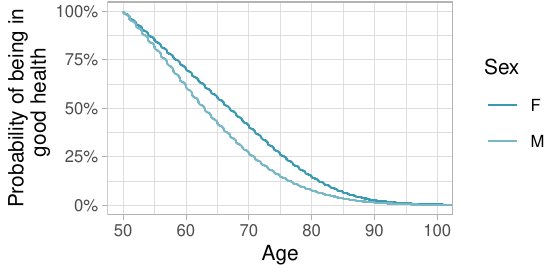}
\caption{\label{fig:health-adj-surv-curve-nocov}Survival curves of being
without disease for the general population aged 50 and up, by sex.}
\end{figure}

\begin{figure}
\centering
\includegraphics{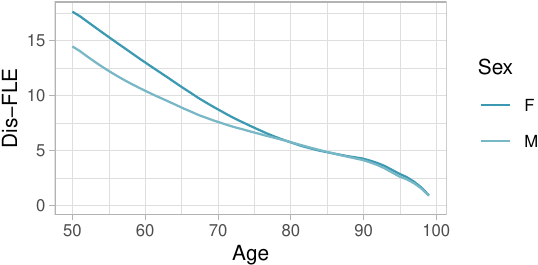}
\caption{\label{fig:health-adj-expect-curve-nocov} Conditional Dis-FLE
adjusted for the whole French population, aged 50 and up, as a function
of age and sex.}
\end{figure}

We estimate Dis-FLE using Kaplan-Meier survival curve estimates on the
data adjusted for the whole population. The data allow us to calculate
the entire survival curve and Dis-FLE for each age. Figures
\ref{fig:health-adj-surv-curve-nocov} and
\ref{fig:health-adj-expect-curve-nocov} show the survival curves and
Dis-FLE with the adjustment for the whole population. Life duration in
good health is significantly larger with than without the whole
population adjustment (see Figures \ref{fig:health-surv-curve-nocov} and
\ref{fig:health-expect-curve-nocov} in the Appendix for the unadjusted
curves).

Overall, Dis-FLE steadily decreases from 50 to about 80, before
stabilizing from about 80 to 90, and continues to decrease thereafter.
Seeing the entire curve reveals an interesting pattern : the sex gap
between Dis-FLE starts at about 5 years at 50, and decreases to 0 at 80.
Dis-FLE for men and women stays essentially the same thereafter. Dis-FLE
without whole population adjustment (Figure
\ref{fig:health-expect-curve-nocov} in the appendix) does display a
proportionally consistent sex gap; therefore, the closing of the sex gap
observed in Figure \ref{fig:health-adj-expect-curve-nocov} is due to the
whole-population adjustment. There is a higher proportion of men than
women who never enter a hospital. Therefore, the adjustment adds more
healthy men than healthy women, thus having a favorable impact for
Dis-FLE for men, relative to women. However, this observation is
difficult to interpret and requires further investigation in future
research. For this reason, we focus on the Cox model for the
hospitalized population only.

To place the proposed Dis-FLE indicator into context of existing health
indicators, we compare it to the closest available indicator, Eurostat's
Healthy Life Years (HLY) for France over the same period (Eurostat
2020). HLY's concept of health is based on a self-evaluation of long
term activity limitation, as measured by GALI of the EU-SILC survey.

HLY represents the expected life duration without long term activity
limitation. This indicator was deliberately chosen to reflect the
overall level of perceived ability, without attempting to identify the
source of type of limitations. This allows it to be simple, and be
widely applied, thus increasing coverage and allowing for comparisons
between countries and over time (Robine 2003).

HLY is also the only comparable health indicator covering France in the
observation period. Another candidate was the HALE indicator from Global
Burden of Disease study for France, but it is not directly comparable to
Dis-FLE-type indicators, as HALE assigns weights to different health
states. Finally, previous articles using these data, (Guibert, Planchet,
and Schwarzinger 2018a, 2018b), focused on similar Dis-FLE type
indicator, but that took into consideration only a small number of
severe diseases, resulting in significantly longer Dis-FLE.

Table \ref{tab:hly-comparaison} compares Dis-FLE adjusted for the whole
French population with HLY at ages 50 and 65. In general, Dis-FLE and
HLY follow expected patterns, decreasing from age 50 to 65 for both
genders. Women consistently exhibit higher Dis-FLE and HLY compared to
men across all ages. However, at age 50, Dis-FLE is significantly lower
than HLY for both genders. Furthermore, the sex gap is more pronounced
in Dis-FLE. At 50, for Dis-FLE the female-male gap is 2 years larger
than for HLY. At 65 the difference between sex gaps is smaller but still
present at about 1 year.

Assuming that the Dis-FLE estimates are indeed representative of the
general population then the difference may be explained by the
difference of perceived activity limitation as measured by GALI and
their clinical state, as well as the exclusion of institutional
households in the EU-SILC survey.

\begin{table}

\caption{\label{tab:hly-comparaison}
            Comparison of Eurostat's HLY at 50 and 65 for France to analogous
            Dis-FLE calculated with the proposed health definition and method.
            HLY value corresponds to the average of HLY from 2010 to 2013.
            The entire Dis-FLE curve can be seen in Figure \ref{fig:health-adj-expect-curve-nocov}.
        }
\centering
\begin{tabular}[t]{rlll}
\toprule
Age & Sex & Dis-FLE & HLY\\
\midrule
50 & Men & 14.5 & 18.8\\
50 & Women & 17.6 & 19.9\\
65 & Men & 8.9 & 9.5\\
65 & Women & 10.8 & 10.2\\
\bottomrule
\end{tabular}
\end{table}

\hypertarget{cox-model-inferences}{%
\subsection{\texorpdfstring{Cox model inferences
\label{sec:cox}}{Cox model inferences }}\label{cox-model-inferences}}

In this section we analyze the data through a Cox model described in
Section \ref{sec:stat_methods}. This model allows us to identify factors
influencing health. Through this analysis, we illustrate the advantages
of using clinical data. Similar analysis would not be possible with
other data sources, either because they lack the necessary information
(covariables) or volume.

We present hazard ratios estimated for this Cox model, that is,
\(e^{\beta_j}\) for the \(j^\text{th}\) variable, rather than the model
coefficient, \(\beta_j\). For non-age-dependent effects we give the
numeric value of the ratio in a table. For terms with age-dependent
effects we show curves of ratios as a function of age.

Overall, the available covariables have a large impact on healthy life
duration, with behavioral risk factors having the largest impact, but
that impact also decreases with age. Following risk factors in
importance is sex, with men experiencing adverse events earlier than
women, even after controlling for covariables. As with risk factors, the
difference becomes smaller for later ages.

In the following sections we examine one-by-one the effects of the risk
factors, but first we want to get an overall idea of just how much the
risk factors influence Dis-FLE.

N.B. : the estimates of Dis-FLE and other quantities do not represent
estimates for the general French population as the adjustment described
in Section \ref{sec:whole_pop_adj} cannot be applied for the Cox model.

\hypertarget{risk-profiles}{%
\subsubsection{Risk profiles}\label{risk-profiles}}

Before delving into the individual impact of each variable, we first
illustrate the collective discriminatory power of the model by examining
survival curves and Dis-FLE for selected risk profiles. As will be seen
later in this section, the presence of risk-increasing behaviors present
(smoking, obesity, and alcohol consumption) is the determinant factor of
Dis-FLE. Therefore, the risk profiles are simply the number of
risk-increasing behaviors present (smoking, obesity, and alcohol
consumption) :

\begin{itemize}
\tightlist
\item
  The ``Lowest'' risk profile, representing individuals without any risk
  factors.
\item
  The ``Intermediate'' risk profile, involving one risk-increasing
  behavior.
\item
  The ``Highest'' risk profile, featuring two risk-increasing behaviors.
\end{itemize}

\begin{figure}
\centering
\includegraphics{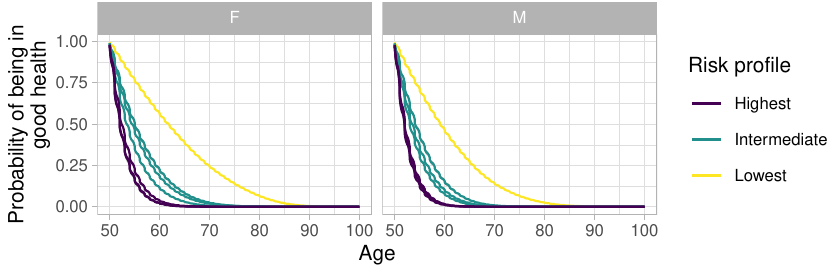}
\caption{\label{fig:cox_f3_profile_surv_curves} Survival curves for
selected risk profiles, by sex, with 95\% confidence intervals.}
\end{figure}

\begin{figure}
\centering
\includegraphics{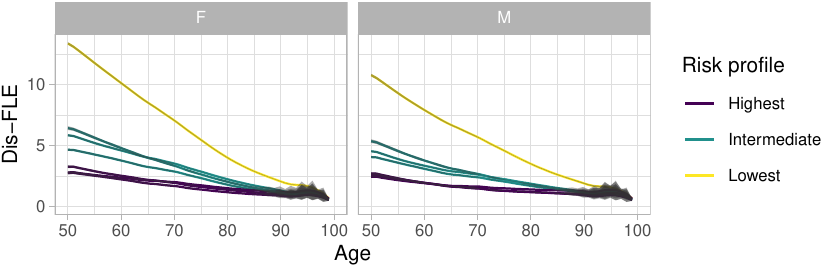}
\caption{\label{fig:cox_f3_profile_surv_time_curves} Conditional
residual expectation for selected risk profiles, by sex, with 95\%
confidence intervals.}
\end{figure}

Figures \ref{fig:cox_f3_profile_surv_curves} and
\ref{fig:cox_f3_profile_surv_time_curves} display survival curves and
\(\text{Dis-FLE}(t)\) estimated by the Cox model for these risk
profiles. There are two curves for the ``Lowest'' risk profile, one for
men and one for women, while the ``Intermediate'' and ``Highest''
profiles each include six curves, one for each combination of sex and
one of the risk factors. Since these risk profiles are just groupings of
covariables, they remain constant for each individual.

The impact of risk behaviors on disease-free life duration is evident,
with a substantial 10-year range in Dis-FLE at age 50 between the lowest
and highest risk profiles. Having at least one risk-increasing behavior
appears to be a key factor, reducing Dis-FLE by approximately 5 years.
In the absence of such behaviors, sex emerges as the determining factor
for Dis-FLE.

It's worth noting that Dis-FLE curves may intersect for men and women in
some risk profiles due to age-dependent coefficients in the Cox model.
Additionally, these figures allow us to isolate the sex gap when other
factors are equal. For instance, in the absence of risk factors at age
50, the sex gap is approximately 2.5 years. However, with the presence
of at least one risk factor, this gap diminishes to less than a year.
This indicates that while behavioral differences contribute to the
Dis-FLE sex gap, they do not entirely explain it.

\hypertarget{sex}{%
\subsubsection{Sex}\label{sex}}

\begin{figure}
\centering
\includegraphics{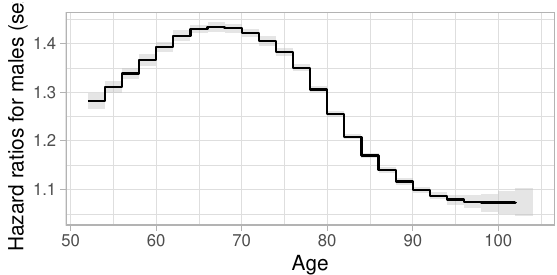}
\caption{\label{fig:cox_f1_tt1_effects_sex} Estimated age-dependent
hazard ratio for sex. Values above 1 increase hazard for males. Gray
areas are 95\% pointwise confidence intervals.}
\end{figure}

We now proceed to inspect the effect of each variable on the
disease-free life duration one by one. We examine age-dependent hazard
ratios. First variable analyzed is the sex of the individual. To take
into account apparent non-proprotionality of hazard functions, the
estimated hazard ratio of sex is allowed to vary with age and is modeled
by a step function. All else being equal men have larger hazard than
women, even when controlling for other covariates, as seen in Figure
\ref{fig:cox_f1_tt1_effects_sex}. This difference is not constant over
time, it starts off at about 30\% excess hazard at 50, and rises
steadily before attaining a maximum of almost 45\% excess hazard at
about 70 years of age. The difference then declines to 5\% at 100 years.

Note that Figure \ref{fig:cox_f3_profile_surv_time_curves} illustrates
the impact of sex on Dis-FLE while keeping other variables constant.
From it, we see that in absence of risk factors Dis-FLE is 2.5 years
lower for men than for women. In presence of at least one risk factor
the difference is less than a year.

\hypertarget{behavioral-risk-factors}{%
\subsubsection{Behavioral risk factors}\label{behavioral-risk-factors}}

We analyzed the effect of three risk factors :

\begin{itemize}
\tightlist
\item
  tobacco consumption,
\item
  alcohol consumption,
\item
  obesity.
\end{itemize}

\begin{figure}
\centering
\includegraphics{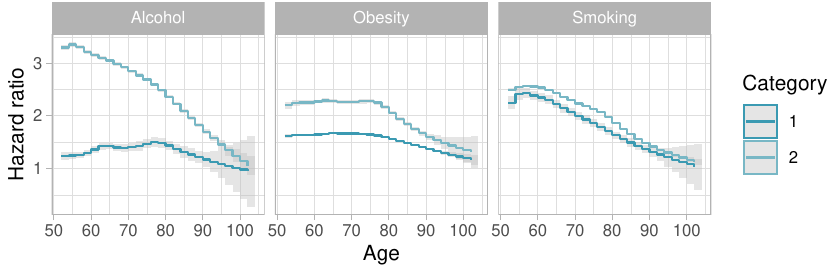}
\caption{\label{fig:cox_f1_tt1_effects_fdr} Estimated age-dependent
hazard ratios for behavioral risk factors. Values above 1 increase
hazard. Gray areas are 95\% pointwise confidence intervals.}
\end{figure}

Each risk factor is grouped into three risk categories, 0, 1, and 2.
Category 0 represents the absence of risk-increasing behavior and is
taken as reference. Figure \ref{fig:cox_f1_tt1_effects_fdr} shows the
age-dependent effects for these risk factors. All risk factors appear to
have large negative impact on outcome. The impact of these risk factors
appears to decrease with age.

Category 2 alcohol abuse has the largest impact on health (although it
also impacts the smallest population compared to other risk factors),
followed by smoking and obesity. The hazard ratios for category 1 risk
factors are substantially smaller. All hazard ratios decrease with age.

\hypertarget{multiple-behavioral-risk-factors}{%
\subsubsection{Multiple behavioral risk
factors}\label{multiple-behavioral-risk-factors}}

In our analysis, we investigated the combined impact of multiple risk
factors. Given the extensive range of possible combinations involving
category 1 and 2 risk factors, we specifically concentrated on the most
prevalent interactions---those among category 2 risk factors.

We find that multiple risk factors increases the overall risk. However,
the marginal increase in risk is less pronounced compared to the risk
associated with each factor independently. This suggests a compensatory
effect when multiple behavioral risk factors coexist. Notably, the
combination of alcohol and smoking exhibits the highest compensatory
ratio, followed by obesity-alcohol and obesity-smoking.

\begin{figure}
\centering
\includegraphics{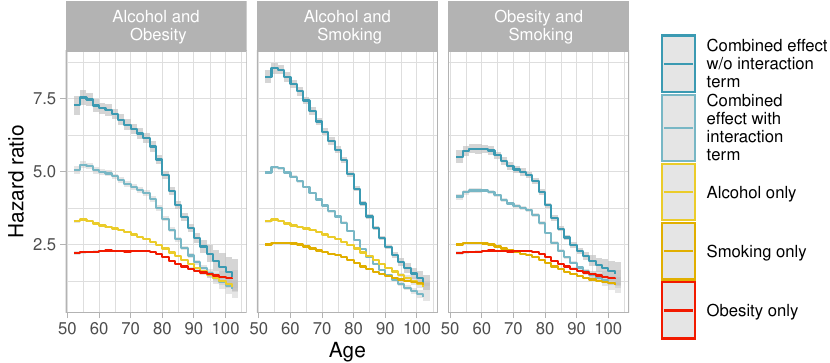}
\caption{\label{fig:cox-f1-tt1-effects-fdr-inter} Estimated
age-dependent hazard ratios for two-way category 2 risk factors
combinations. Each panel shows the interplay between two risk factors.
For each panel, the main age-dependent effect is shown for the risk
factors, and the combined effect with and without interaction are
displayed. The combined effect without interaction is simply the product
of the hazard ratios of the main effect. The combined effect with
interactions is the product of the main effects and the interaction
term.. Values above 1 increase hazard. Gray areas are 95\% pointwise
confidence intervals.}
\end{figure}

Figure \ref{fig:cox-f1-tt1-effects-fdr-inter} visually represents the
distinctions between :

\begin{itemize}
\tightlist
\item
  the main effects,
\item
  the naive combined effect of two risk factors (calculated by
  multiplying the hazard ratios of the main effects without considering
  the interaction term),
\item
  and the estimated effect that accounts for the interaction term.
\end{itemize}

For all three combinations of risk factors the combined effect with
interaction is lower than without it. These observations shed light on
the nuanced interplay of risk factors and their collective influence on
the overall hazard.

\hypertarget{behavioral-risk-factors-conditional-on-sex}{%
\subsubsection{Behavioral risk factors conditional on
sex}\label{behavioral-risk-factors-conditional-on-sex}}

We measure if risk factors impact men and women differently. To simplify
the model we model this difference as an offset for males. Table
\ref{tab:cox-f3-coefs-fdr-inter} gives the hazard ratios for the
interaction terms between sex and behavioral risk factors. These ratios
can be interpreted as additional burden of these risk factors on men,
relative to women.

Overall men appear to be slightly less sensitive to the presence of
behavioral risk factors. This explains in part the reason for decrease
in the Dis-FLE sex gap in presence of risk factors, as seen in Figure
\ref{fig:cox_f3_profile_surv_time_curves}.

We focus on category 2 behavioral risk factors because category 1 are
rare or without substantial male-female differences. Category 2 alcohol
consumption, has a substantially stronger impact on women, with women
suffering additional 12\% of hazard. Obesity also impacts women
stronger, by about 7\%; While men's health is slightly more sensitive to
smoking.

\begin{table}

\caption{\label{tab:cox-f3-coefs-fdr-inter}
            Hazard ratios for additional risk for men from behavioral
            risk factors, with associated standard errors and p-values.
            Only category 2 risk factors are shown.
            }
\centering
\begin{tabular}[t]{llll}
\toprule
Risk factor (Cat. 2) & Hazard ratio & Std. error & p-value\\
\midrule
Obesity & 0.934 & 0.010 & 0.000\\
Alcohol & 0.882 & 0.007 & 0.000\\
Smoking & 1.010 & 0.004 & 0.022\\
\bottomrule
\end{tabular}
\end{table}

\hypertarget{geographical}{%
\subsubsection{Geographical}\label{geographical}}

\begin{figure}
\centering
\includegraphics{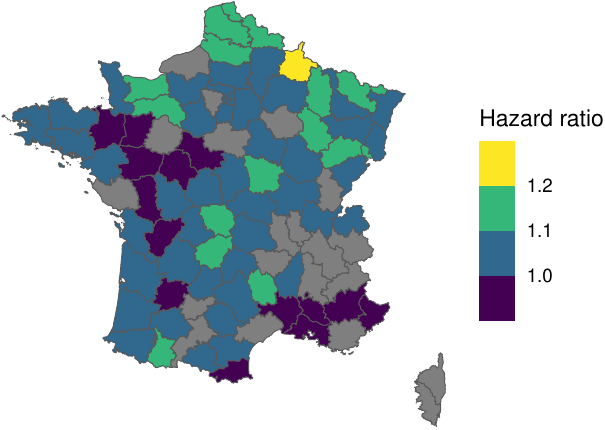}
\caption{\label{fig:cox-f3-coefs-dep} Estimated hazard ratios for
departments of residence. Values binned. Values above 1 increase hazard
relative to residents of department 78. Non-significant values (p-value
\textgreater{} 0.05) are greyed out.}
\end{figure}

Figure \ref{fig:cox-f3-coefs-dep} gives the hazard ratios relative to
the Yvelines department (78). This reference was chosen because it is
the Île-de-France region, while not being Paris itself.

Northern departments have a markedly higher hazard rate, even after
controlling for other covariates. South-east, and eastern departments on
the other hand appear to have the inverse effect. Both these facts are
in accord with previous literature. In the rest of the territory the
effects appear to be more local.

\begin{figure}
\centering
\includegraphics{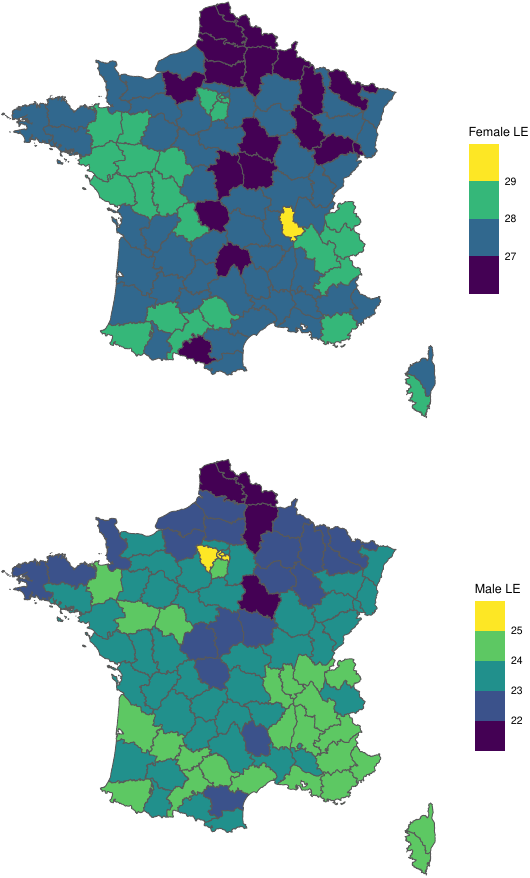}
\caption{\label{fig:insee-le-dep} Life expectany at 60, by sex, in
France in 2023. Values binned.}
\end{figure}

To put these results in context Figure \ref{fig:insee-le-dep} provides a
map of life expectancies at 60 by sex and by department (INSEE 2023).
Overall, we observe similar trends. The similarity suggests that the
geographic location is an independent predictor of life expectancy and
Dis-FLE.

In and of itself it is hard to interpret this result, as may not
necessarily reflect the impact of local environment on health, but
instead reflect the level of access to healthcare, as discussed when
introducing this approach. Further work is necessary to explain these
differences. A first step would be including more information on the
departments themselves, e.g., population, population density, GDP,
median income, etc.

The variables ``Education'' and ``Immigration'' indicate the level of
education and immigration in the commune of residence. Table
\ref{tab:cox-f3-coefs-immi-dip} gives the obtained hazard ratios for
these variables. Surprisingly, the level of education and immigration in
the commune of residence appear to increase the hazard. The effect is
minor compared to other risk factors, but nonetheless significant. This
result is also hard to interpret on it own as there is a level of
indirection between the individual and the commune of residence.

\begin{longtable}{rrrr}
\caption{
\label{tab:cox-f3-coefs-immi-dip}Cox model coefficients for the education and immigration levels in the commune of residence.
} \\ 
\toprule
Quartile & Hazard ratio & Std. error & p-value \\ 
\midrule
\multicolumn{4}{l}{Immigration} \\ 
\midrule
1 & 1.003 & 0.002 & 0.250 \\ 
2 & 1.006 & 0.002 & 0.004 \\ 
3 & 1.008 & 0.002 & 0.001 \\ 
\midrule
\multicolumn{4}{l}{Education} \\ 
\midrule
1 & 1.029 & 0.002 & 0.000 \\ 
2 & 1.051 & 0.002 & 0.000 \\ 
3 & 1.071 & 0.002 & 0.000 \\ 
\bottomrule
\end{longtable}

\hypertarget{conclusion}{%
\section{\texorpdfstring{Conclusion
\label{sec:discuss}}{Conclusion }}\label{conclusion}}

We propose the use of clinical data to construct health indicators. The
use of clinical data opens up a hitherto unused source of information
and makes rich analysis possible, some of which we present in this
paper.

This work provides a methodological blueprint for calculating health
indicators based on clinical datasets. The implications of our research
extend beyond the French context, with potential applications in other
countries and healthcare systems. Specifically, our methodology is not
confined to large clinical datasets and can be applied at smaller
scales, such as hospital cohorts, in France or elsewhere. However, when
considering entire populations, accessing national hospitalization
datasets to calculate nationally representative health indicators can be
exceedingly challenging. We hope that this work provides a precedent
that will encourage and facilitate similar efforts in the future.

Although clinical data impose a diagnosis centric vision, rather than
outcome based one that may be provided by health oriented survey
instruments, it does provide with a clear outline of the health state
over the lifetime of the patient. When combined with the large volume
data available, this results in pertinent indicators on a population
level. Indeed, as the comparison with HLY shows, Dis-FLE with the
adjustment for the whole population displays similar trends, although
with a wider sex gap.

In the absence of standardized practice to define health from clinical
data it is difficult to construct comparable health indicators. We
sidestep the issue by focusing on a simple definition of being
disease-free. A more complex indicator would take into account the
entire health trajectory, but would be difficult to analyze, something
that could be treated in further work. Instead, our focus on simple
trajectories combined with large amount of data available allowed us to
exhaustively analyze the impact of available covariables. In doing so we
illustrate the kind of analysis we believe can be made possible by using
clinical data. We apply the proposed methods to the French PMSI
database, and analyze the health status of population aged 50 and up
from 2010 to 2013. We summarize the results of the analysis in terms of
Dis-FLE based on 36 severe conditions and hazard ratios of the
corresponding Cox model.

For the population studied the Dis-FLE at 50 years is 10 years for women
and 7.5 for men. Dis-FLE is strongly influenced by the covariables
available, indeed Dis-FLE can range from 2.5 to 12.5 for women and from
2.5 to 10 for men when conditioned on covariates.

The most important determinant of Dis-FLE are the behavioral factors, in
order of importance : alcohol consumption, tobacco use and obesity. Each
of these have hazard ratios exceeding 2 for all ages before 80. Alcohol
consumption has hazard ratio larger than 3 before 60 years.
Interestingly, all age-dependent effects decrease with age after 60.

Sex also has a large influence with a hazard ratio above 1.2 before 80,
and as large as 1.4 at about 70. Also, the effect of behavioral risk
factors was found to differ by sex, with alcohol consumption and obesity
having a stronger effect on women, and smoking having a stronger effect
on men. Other factors influence Dis-FLE, but have a weaker effect.

The Cox model analyzed in this paper is the simplest model that still
allows us to illustrate the richness of underlying data. There are
however many possible improvements to it. For example, the model
analyzed does not take into account calendar time. This is in contrast
to most indicators where the ability to follow them over time is vital.
A natural extension of the model would be to take in account calendar
time by including it as an age-dependent covariate. Other possible
extensions include making effects not only depend on age, but also on
calendar time, therefore modeling possible improvements of treatment of
behavioral factors.

The trajectories analyzed are based on a specific definition of health,
or more specifically disease-free. This definition is based on previous
work using this dataset, and is conceptually coherent with other
indicators. However, it lacks direct comparables, making its usefulness
as an indicator limited for now. Further work may help identify a
definition of disability closer aligned with other indicators, such as
GALI.

More fundamentally, the concept of health used introduces an artificial
dichotomy between good and ill health. Using the same data it should be
possible to define more realistic individual trajectories, for example
by assigning each disease a weight. Using this approach we can define
individual level health-weighted indicator, extending the flexible
approach to other indicators such as HALE. In this context the use of
clinical data would also simply methodology as many problems plaguing
HALE estimates are resolved by these data, as for example comorbidity
and the nuance between incidence and prevalence.

Such an approach would make both the definition of health trajectories
and their analysis significantly more complex. We, believe however that
that would be a natural next step in using clinical data as data-source
for health indicators.

Beyond considerations of health concept used, the use of clinical data
requires additional assumptions and adjustment procedures to produce
nationally-representative indicators. A simple adjustment procedure was
introduced and used to calculate Dis-FLE for the general French
population. However, we believe that this procedure could be improved by
using more granular data and, under additional assumptions, extended to
the estimates provided by the Cox model.

Should our methodology and findings prove useful and robust, future work
could delve into the development of a definition of health, that is
based on clinical data that explicitly targets GALI or other relevant
health indicators, potentially drawing upon detailed assessments of
activities of daily living (ADLs). Such an endeavor could enhance the
accuracy and sensitivity of our understanding of disability and its
implications for individual and population health.

The Cox model was the tool of choice for this analysis. However, the
large volume of data combined with the need to explicitly define the
model matrix required a large amount of computer memory to do the
necessary computations. The use of other machine learning algorithms may
provide a more efficient means to analyze this dataset.

\appendix

\hypertarget{exclusion-criteria}{%
\section{Exclusion criteria}\label{exclusion-criteria}}

Table \ref{tab:filters-schwarz} gives the exclusion criteria applied to
the dataset. Translated and adapted from Table 1 in Schwarzinger (2018).

\begin{longtable}[t]{>{\raggedright\arraybackslash}p{20em}|l|l|l}
\caption{\label{tab:filters-schwarz}Exclusion criteria and impact on number of patients.}\\
\hline
Criteria & Years & Pop. size & \% of total pop.\\
\hline
\cellcolor{gray}{\textbf{\textbf{Hospitalized population, aged 50 and up}}} & \cellcolor{gray}{\textbf{\textbf{2008-2013}}} & \cellcolor{gray}{\textbf{\textbf{18 440 022}}} & \cellcolor{gray}{\textbf{\textbf{100.0\%}}}\\
\hline
\cellcolor{gray}{\textbf{\textbf{Exclusion criterion 1: severe conditions under study observed in 2008-2009}}} & \cellcolor{gray}{\textbf{\textbf{2008,2009}}} & \cellcolor{gray}{\textbf{\textbf{4 730 651}}} & \cellcolor{gray}{\textbf{\textbf{25.7\%}}}\\
\hline
Alzheimer's disease (n=1) &  & 508 575 & 2.8\%\\
\hline
Other severe conditions (n=34) &  & 4 554 010 & 24.7\%\\
\hline
Total loss of autonomy, coginitive or physical (n=2) &  & 205 681 & 1.1\%\\
\hline
Death observed in a hospital &  & 572 454 & 3.1\%\\
\hline
Death outside of hospital (imputed from other data) &  & 272 742 & 1.5\%\\
\hline
\cellcolor{gray}{\textbf{\textbf{Exclusion criterion 2 : other conditions not covered by dependence insurance}}} & \cellcolor{gray}{\textbf{\textbf{}}} & \cellcolor{gray}{\textbf{\textbf{914 595}}} & \cellcolor{gray}{\textbf{\textbf{5.0\%}}}\\
\hline
\textbf{Major neurological disorder} & \textbf{2008,2009} & \textbf{} & \textbf{}\\
\hline
Paralisis &  & 197 096 & 1.1\%\\
\hline
Coma &  & 97 476 & 0.5\%\\
\hline
\textbf{Transplan recepients (of an organ or of bone marrow)} & \textbf{2008,2009} & \textbf{36 593} & \textbf{0.2\%}\\
\hline
\textbf{Birth defects and genetic disorders} & \textbf{2008-2013} & \textbf{} & \textbf{}\\
\hline
Birth defect or Chromosome abnormality (including trisomy 21) &  & 272 887 & 1.5\%\\
\hline
Primary immunodeficiency &  & 37 101 & 0.2\%\\
\hline
Thalassemia, Sickle cell disease and other blood disorders &  & 16 200 & 0.1\%\\
\hline
Haemophilia and other bleeding disorders &  & 16 600 & 0.1\%\\
\hline
Inborn errors of metabolism (including haemochromatosis and cystic fibrosis) &  & 210 176 & 1.1\%\\
\hline
Cerebral palsy and genetic neuromuscular disorders (including myopathy) &  & 70 587 & 0.4\%\\
\hline
Other genetic disorders (including Alport syndrome) &  & 2 252 & 0.0\%\\
\hline
\textbf{Infections} & \textbf{2008-2013} & \textbf{} & \textbf{}\\
\hline
Human immunodeficiency viruses (HIV) &  & 43 734 & 0.2\%\\
\hline
Infectious diseases (including tuberculosis and encephalitis) &  & 49 524 & 0.3\%\\
\hline
\textbf{Mental disorders} & \textbf{2008-2013} & \textbf{} & \textbf{}\\
\hline
Schizophrenia and others delusional disorders &  & 175 527 & 1.0\%\\
\hline
Intellectual disability &  & 44 936 & 0.2\%\\
\hline
\cellcolor{gray}{\textbf{\textbf{Hospitalized population, aged 50 and up in good health on the 1st of January 2010 (selected after exclusion criteria 1 and 2)}}} & \cellcolor{gray}{\textbf{\textbf{2010-2013}}} & \cellcolor{gray}{\textbf{\textbf{13 170 355}}} & \cellcolor{gray}{\textbf{\textbf{71.4\%}}}\\
\hline
\textbf{Data preparation for analysis} & \textbf{2008,2009} & \textbf{2 559 726} & \textbf{13.9\%}\\
\hline
Censored before 2010/01/01 &  & 2 012 815 & 10.9\%\\
\hline
End of observation period ends before 50 &  & 896 111 & 4.9\%\\
\hline
\cellcolor{gray}{\textbf{\textbf{Population included in study}}} & \cellcolor{gray}{\textbf{\textbf{2010-2013}}} & \cellcolor{gray}{\textbf{\textbf{10 610 629}}} & \cellcolor{gray}{\textbf{\textbf{57.5\%}}}\\
\hline
\end{longtable}

\newpage

\hypertarget{sex-specific-survival-curves-without-adjustment}{%
\section{Sex-specific survival curves without
adjustment}\label{sex-specific-survival-curves-without-adjustment}}

\begin{figure}
\centering
\includegraphics{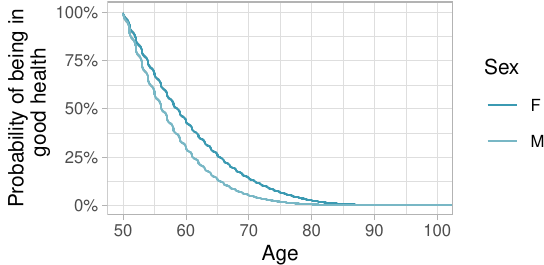}
\caption{\label{fig:health-surv-curve-nocov}Survival curves of being in
good health, by sex.}
\end{figure}

\begin{figure}
\centering
\includegraphics{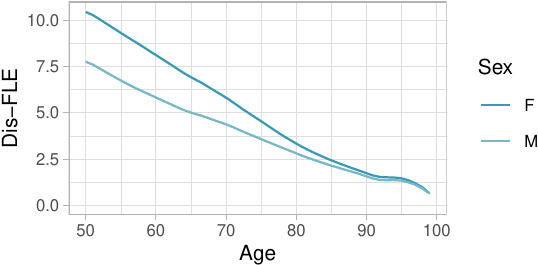}
\caption{\label{fig:health-expect-curve-nocov}Conditional Dis-FLE as a
function of age and sex.}
\end{figure}

Figure \ref{fig:health-surv-curve-nocov} shows the sex specific survival
curves without adjustment for the whole population. Unsurprisingly women
spend longer in healthy state than men. The oscillations in the curves
are due to rounding in anonymized dates. Figure
\ref{fig:health-expect-curve-nocov} shows the corresponding
\(\text{Dis-FLE}(t)\).

\newpage

\hypertarget{model-diagnostics}{%
\section{Model diagnostics}\label{model-diagnostics}}

The cox model used is fit on 60\% of the available data. The remaining
40\% are reserved to perform model diagnostics presented in this
section.

The C-statistic calculated on the test set is 59.91\%.

\begin{figure}
\centering
\includegraphics{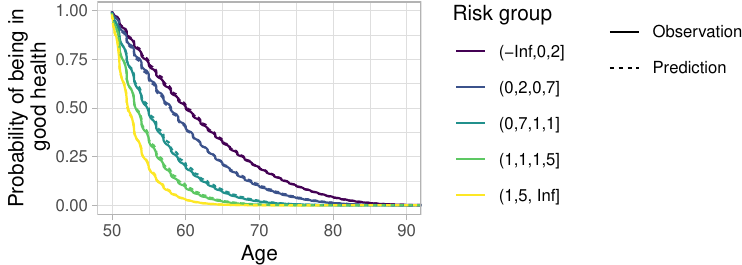}
\caption{\label{fig:compare_surv_test_risk_group} Comparaison between
observed and estimated survival functions for the test dataset, grouped
by linear predictor.}
\end{figure}

To evaluate the quality of the fit on the test dataset, we calculate the
linear predictor, i.e., \(\log(\text{Hazard ratio})\), for every
individual. Individuals are then classified into classes based on the
calculated value. The distribution of linear predictors is clustered
around few values. This is due to the fact that the influence of sex and
the presence of risk factors essentially determines risk, with all other
variables essentially only adding noise. The lowest interval :
\((-\infty, 0.2]\) covers essentially only women without any risk
factors; \((0.2, 0.7]\) covers men without any risk factors;
\((0.7; 1.1]\) covers mostly persons with obesity, \((1.1, 1,5]\) covers
mostly smokers, and \((1,5, \infty]\) covers persons with alcohol
consumption, or multiple risk factors. Finally, Figure
\ref{fig:compare_surv_test_risk_group} compares the observed survival
curves for each of these classes with the predicted survival curve.

One problem with this approach is that the model includes age-dependent
coefficients. This means that risk score for each individual changes
over time, making it impossible to attribute a constant score to each
individual. However, since the observation period is four years and the
time grid for age-dependent coefficient is two years, each individual
may at most have 3 unique risk values, and most have only 1. When
multiple risk values are present, they are close to each other. For the
calculation above, we use the average of predicted linear prediction
scores.

\newpage

\hypertarget{all-cox-model-coefficients}{%
\section{All Cox model coefficients}\label{all-cox-model-coefficients}}

\begin{longtable}{lrrr}
\caption{
\label{tab:all_cox_coefs}All Cox model coefficient with associated standard errors and p-values.
} \\ 
\toprule
Term & Hazard ratio & Std. err. & p-value \\ 
\midrule
Obesity, cat. 1 & 1.62 & 0.01 & 0.00 \\ 
Obesity, cat. 2 & 2.20 & 0.02 & 0.00 \\ 
Alcohol, cat. 1 & 1.24 & 0.03 & 0.00 \\ 
Alcohol, cat. 2 & 3.30 & 0.01 & 0.00 \\ 
Smoking, cat. 1 & 2.25 & 0.04 & 0.00 \\ 
Smoking, cat. 2 & 2.50 & 0.01 & 0.00 \\ 
Sex : M & 1.28 & 0.01 & 0.00 \\ 
Department of residence : 02 & 1.07 & 0.01 & 0.00 \\ 
Department of residence : 03 & 1.05 & 0.01 & 0.00 \\ 
Department of residence : 04 & 0.95 & 0.01 & 0.00 \\ 
Department of residence : 05 & 1.02 & 0.02 & 0.16 \\ 
Department of residence : 06 & 0.91 & 0.01 & 0.00 \\ 
Department of residence : 07 & 1.03 & 0.01 & 0.01 \\ 
Department of residence : 08 & 1.20 & 0.01 & 0.00 \\ 
Department of residence : 09 & 1.06 & 0.01 & 0.00 \\ 
Department of residence : 10 & 1.00 & 0.01 & 0.83 \\ 
Department of residence : 11 & 1.07 & 0.01 & 0.00 \\ 
Department of residence : 12 & 1.04 & 0.01 & 0.00 \\ 
Department of residence : 13 & 0.95 & 0.01 & 0.00 \\ 
Department of residence : 14 & 1.09 & 0.01 & 0.00 \\ 
Department of residence : 15 & 1.03 & 0.01 & 0.02 \\ 
Department of residence : 16 & 0.96 & 0.01 & 0.00 \\ 
Department of residence : 17 & 1.06 & 0.01 & 0.00 \\ 
Department of residence : 18 & 1.08 & 0.01 & 0.00 \\ 
Department of residence : 19 & 1.10 & 0.01 & 0.00 \\ 
Department of residence : 21 & 1.03 & 0.01 & 0.00 \\ 
Department of residence : 22 & 1.08 & 0.01 & 0.00 \\ 
Department of residence : 23 & 1.09 & 0.01 & 0.00 \\ 
Department of residence : 24 & 1.05 & 0.01 & 0.00 \\ 
Department of residence : 25 & 1.06 & 0.01 & 0.00 \\ 
Department of residence : 26 & 1.01 & 0.01 & 0.59 \\ 
Department of residence : 27 & 1.04 & 0.01 & 0.00 \\ 
Department of residence : 28 & 1.04 & 0.01 & 0.00 \\ 
Department of residence : 29 & 1.03 & 0.01 & 0.00 \\ 
Department of residence : 2A & 1.03 & 0.01 & 0.06 \\ 
Department of residence : 2B & 0.99 & 0.01 & 0.46 \\ 
Department of residence : 30 & 0.98 & 0.01 & 0.01 \\ 
Department of residence : 31 & 1.01 & 0.01 & 0.33 \\ 
Department of residence : 32 & 1.06 & 0.01 & 0.00 \\ 
Department of residence : 33 & 1.02 & 0.01 & 0.04 \\ 
Department of residence : 34 & 1.00 & 0.01 & 0.97 \\ 
Department of residence : 35 & 0.98 & 0.01 & 0.05 \\ 
Department of residence : 36 & 1.03 & 0.01 & 0.01 \\ 
Department of residence : 37 & 0.98 & 0.01 & 0.05 \\ 
Department of residence : 38 & 1.00 & 0.01 & 0.64 \\ 
Department of residence : 39 & 1.02 & 0.01 & 0.14 \\ 
Department of residence : 40 & 1.03 & 0.01 & 0.01 \\ 
Department of residence : 41 & 0.98 & 0.01 & 0.04 \\ 
Department of residence : 42 & 0.98 & 0.01 & 0.06 \\ 
Department of residence : 43 & 1.01 & 0.01 & 0.27 \\ 
Department of residence : 44 & 1.03 & 0.01 & 0.00 \\ 
Department of residence : 45 & 1.01 & 0.01 & 0.14 \\ 
Department of residence : 46 & 1.04 & 0.01 & 0.00 \\ 
Department of residence : 47 & 0.97 & 0.01 & 0.00 \\ 
Department of residence : 48 & 1.11 & 0.02 & 0.00 \\ 
Department of residence : 49 & 0.94 & 0.01 & 0.00 \\ 
Department of residence : 50 & 1.07 & 0.01 & 0.00 \\ 
Department of residence : 51 & 1.03 & 0.01 & 0.00 \\ 
Department of residence : 52 & 1.13 & 0.01 & 0.00 \\ 
Department of residence : 53 & 0.93 & 0.01 & 0.00 \\ 
Department of residence : 54 & 1.05 & 0.01 & 0.00 \\ 
Department of residence : 55 & 1.13 & 0.01 & 0.00 \\ 
Department of residence : 56 & 1.07 & 0.01 & 0.00 \\ 
Department of residence : 57 & 1.14 & 0.01 & 0.00 \\ 
Department of residence : 58 & 1.10 & 0.01 & 0.00 \\ 
Department of residence : 59 & 1.10 & 0.01 & 0.00 \\ 
Department of residence : 60 & 1.04 & 0.01 & 0.00 \\ 
Department of residence : 61 & 1.09 & 0.01 & 0.00 \\ 
Department of residence : 62 & 1.12 & 0.01 & 0.00 \\ 
Department of residence : 63 & 1.08 & 0.01 & 0.00 \\ 
Department of residence : 64 & 1.08 & 0.01 & 0.00 \\ 
Department of residence : 65 & 1.09 & 0.01 & 0.00 \\ 
Department of residence : 66 & 0.96 & 0.01 & 0.00 \\ 
Department of residence : 67 & 1.05 & 0.01 & 0.00 \\ 
Department of residence : 68 & 1.04 & 0.01 & 0.00 \\ 
Department of residence : 69 & 1.02 & 0.01 & 0.06 \\ 
Department of residence : 70 & 1.09 & 0.01 & 0.00 \\ 
Department of residence : 71 & 1.02 & 0.01 & 0.02 \\ 
Department of residence : 72 & 1.00 & 0.01 & 0.86 \\ 
Department of residence : 73 & 1.02 & 0.01 & 0.07 \\ 
Department of residence : 74 & 1.02 & 0.01 & 0.02 \\ 
Department of residence : 75 & 1.03 & 0.01 & 0.00 \\ 
Department of residence : 76 & 1.01 & 0.01 & 0.11 \\ 
Department of residence : 77 & 1.05 & 0.01 & 0.00 \\ 
Department of residence : 78 & 0.98 & 0.01 & 0.08 \\ 
Department of residence : 79 & 0.96 & 0.01 & 0.00 \\ 
Department of residence : 80 & 1.08 & 0.01 & 0.00 \\ 
Department of residence : 81 & 1.06 & 0.01 & 0.00 \\ 
Department of residence : 82 & 0.98 & 0.01 & 0.05 \\ 
Department of residence : 83 & 0.99 & 0.01 & 0.20 \\ 
Department of residence : 84 & 0.94 & 0.01 & 0.00 \\ 
Department of residence : 85 & 1.00 & 0.01 & 0.91 \\ 
Department of residence : 86 & 1.03 & 0.01 & 0.00 \\ 
Department of residence : 87 & 1.07 & 0.01 & 0.00 \\ 
Department of residence : 88 & 1.03 & 0.01 & 0.00 \\ 
Department of residence : 89 & 1.07 & 0.01 & 0.00 \\ 
Department of residence : 90 & 1.09 & 0.02 & 0.00 \\ 
Department of residence : 91 & 1.02 & 0.01 & 0.01 \\ 
Department of residence : 92 & 1.03 & 0.01 & 0.00 \\ 
Department of residence : 93 & 1.07 & 0.01 & 0.00 \\ 
Department of residence : 94 & 1.04 & 0.01 & 0.00 \\ 
Department of residence : 95 & 1.04 & 0.01 & 0.00 \\ 
Immigration : Q1 & 1.00 & 0.00 & 0.25 \\ 
Immigration : Q2 & 1.01 & 0.00 & 0.00 \\ 
Immigration : Q3 & 1.01 & 0.00 & 0.00 \\ 
Education : Q1 & 1.03 & 0.00 & 0.00 \\ 
Education : Q2 & 1.05 & 0.00 & 0.00 \\ 
Education : Q3 & 1.07 & 0.00 & 0.00 \\ 
Obesity, cat. 1 * Alcohol, cat. 1 & 0.84 & 0.03 & 0.00 \\ 
Obesity, cat. 2 * Alcohol, cat. 1 & 0.88 & 0.06 & 0.03 \\ 
Obesity, cat. 1 * Alcohol, cat. 2 & 0.76 & 0.01 & 0.00 \\ 
Obesity, cat. 2 * Alcohol, cat. 2 & 0.69 & 0.02 & 0.00 \\ 
Obesity, cat. 1 * Smoking, cat. 1 & 0.68 & 0.02 & 0.00 \\ 
Obesity, cat. 2 * Smoking, cat. 1 & 0.64 & 0.05 & 0.00 \\ 
Obesity, cat. 1 * Smoking, cat. 2 & 0.77 & 0.01 & 0.00 \\ 
Obesity, cat. 2 * Smoking, cat. 2 & 0.75 & 0.01 & 0.00 \\ 
Obesity, cat. 1 * Sex : M & 0.98 & 0.00 & 0.00 \\ 
Obesity, cat. 2 * Sex : M & 0.93 & 0.01 & 0.00 \\ 
Alcohol, cat. 1 * Smoking, cat. 1 & 0.78 & 0.04 & 0.00 \\ 
Alcohol, cat. 2 * Smoking, cat. 1 & 0.53 & 0.03 & 0.00 \\ 
Alcohol, cat. 1 * Smoking, cat. 2 & 0.67 & 0.02 & 0.00 \\ 
Alcohol, cat. 2 * Smoking, cat. 2 & 0.60 & 0.01 & 0.00 \\ 
Alcohol, cat. 1 * Sex : M & 1.01 & 0.02 & 0.63 \\ 
Alcohol, cat. 2 * Sex : M & 0.88 & 0.01 & 0.00 \\ 
Smoking, cat. 1 * Sex : M & 0.92 & 0.02 & 0.00 \\ 
Smoking, cat. 2 * Sex : M & 1.01 & 0.00 & 0.02 \\ 
Obesity, cat. 1 * Spline(age): knot 1 & 1.01 & 0.01 & 0.48 \\ 
Obesity, cat. 2 * Spline(age): knot 1 & 1.02 & 0.03 & 0.46 \\ 
Obesity, cat. 1 * Spline(age): knot 2 & 1.01 & 0.01 & 0.49 \\ 
Obesity, cat. 2 * Spline(age): knot 2 & 1.02 & 0.03 & 0.34 \\ 
Obesity, cat. 1 * Spline(age): knot 3 & 1.03 & 0.01 & 0.05 \\ 
Obesity, cat. 2 * Spline(age): knot 3 & 1.04 & 0.03 & 0.12 \\ 
Obesity, cat. 1 * Spline(age): knot 4 & 1.03 & 0.01 & 0.02 \\ 
Obesity, cat. 2 * Spline(age): knot 4 & 1.03 & 0.02 & 0.25 \\ 
Obesity, cat. 1 * Spline(age): knot 5 & 1.03 & 0.01 & 0.05 \\ 
Obesity, cat. 2 * Spline(age): knot 5 & 1.03 & 0.03 & 0.19 \\ 
Obesity, cat. 1 * Spline(age): knot 6 & 1.01 & 0.01 & 0.66 \\ 
Obesity, cat. 2 * Spline(age): knot 6 & 1.02 & 0.03 & 0.35 \\ 
Obesity, cat. 1 * Spline(age): knot 7 & 0.94 & 0.01 & 0.00 \\ 
Obesity, cat. 2 * Spline(age): knot 7 & 0.88 & 0.03 & 0.00 \\ 
Obesity, cat. 1 * Spline(age): knot 8 & 0.70 & 0.05 & 0.00 \\ 
Obesity, cat. 2 * Spline(age): knot 8 & 0.58 & 0.16 & 0.00 \\ 
Alcohol, cat. 1 * Spline(age): knot 1 & 1.01 & 0.03 & 0.76 \\ 
Alcohol, cat. 2 * Spline(age): knot 1 & 1.02 & 0.01 & 0.25 \\ 
Alcohol, cat. 1 * Spline(age): knot 2 & 1.04 & 0.03 & 0.18 \\ 
Alcohol, cat. 2 * Spline(age): knot 2 & 0.98 & 0.01 & 0.10 \\ 
Alcohol, cat. 1 * Spline(age): knot 3 & 1.15 & 0.04 & 0.00 \\ 
Alcohol, cat. 2 * Spline(age): knot 3 & 0.94 & 0.02 & 0.00 \\ 
Alcohol, cat. 1 * Spline(age): knot 4 & 1.13 & 0.04 & 0.00 \\ 
Alcohol, cat. 2 * Spline(age): knot 4 & 0.91 & 0.01 & 0.00 \\ 
Alcohol, cat. 1 * Spline(age): knot 5 & 1.16 & 0.04 & 0.00 \\ 
Alcohol, cat. 2 * Spline(age): knot 5 & 0.84 & 0.02 & 0.00 \\ 
Alcohol, cat. 1 * Spline(age): knot 6 & 1.21 & 0.05 & 0.00 \\ 
Alcohol, cat. 2 * Spline(age): knot 6 & 0.79 & 0.02 & 0.00 \\ 
Alcohol, cat. 1 * Spline(age): knot 7 & 1.11 & 0.05 & 0.03 \\ 
Alcohol, cat. 2 * Spline(age): knot 7 & 0.68 & 0.02 & 0.00 \\ 
Alcohol, cat. 1 * Spline(age): knot 8 & 0.74 & 0.41 & 0.47 \\ 
Alcohol, cat. 2 * Spline(age): knot 8 & 0.29 & 0.10 & 0.00 \\ 
Smoking, cat. 1 * Spline(age): knot 1 & 1.07 & 0.05 & 0.14 \\ 
Smoking, cat. 2 * Spline(age): knot 1 & 1.02 & 0.01 & 0.11 \\ 
Smoking, cat. 1 * Spline(age): knot 2 & 1.06 & 0.05 & 0.20 \\ 
Smoking, cat. 2 * Spline(age): knot 2 & 1.02 & 0.01 & 0.03 \\ 
Smoking, cat. 1 * Spline(age): knot 3 & 1.02 & 0.05 & 0.60 \\ 
Smoking, cat. 2 * Spline(age): knot 3 & 1.00 & 0.01 & 0.96 \\ 
Smoking, cat. 1 * Spline(age): knot 4 & 0.95 & 0.05 & 0.28 \\ 
Smoking, cat. 2 * Spline(age): knot 4 & 0.95 & 0.01 & 0.00 \\ 
Smoking, cat. 1 * Spline(age): knot 5 & 0.86 & 0.05 & 0.00 \\ 
Smoking, cat. 2 * Spline(age): knot 5 & 0.87 & 0.01 & 0.00 \\ 
Smoking, cat. 1 * Spline(age): knot 6 & 0.79 & 0.05 & 0.00 \\ 
Smoking, cat. 2 * Spline(age): knot 6 & 0.83 & 0.01 & 0.00 \\ 
Smoking, cat. 1 * Spline(age): knot 7 & 0.69 & 0.05 & 0.00 \\ 
Smoking, cat. 2 * Spline(age): knot 7 & 0.70 & 0.01 & 0.00 \\ 
Smoking, cat. 1 * Spline(age): knot 8 & 0.44 & 0.26 & 0.00 \\ 
Smoking, cat. 2 * Spline(age): knot 8 & 0.44 & 0.03 & 0.00 \\ 
Sex : M * Spline(age): knot 1 & 1.02 & 0.01 & 0.01 \\ 
Sex : M * Spline(age): knot 2 & 1.07 & 0.01 & 0.00 \\ 
Sex : M * Spline(age): knot 3 & 1.10 & 0.01 & 0.00 \\ 
Sex : M * Spline(age): knot 4 & 1.12 & 0.01 & 0.00 \\ 
Sex : M * Spline(age): knot 5 & 1.10 & 0.01 & 0.00 \\ 
Sex : M * Spline(age): knot 6 & 1.05 & 0.01 & 0.00 \\ 
Sex : M * Spline(age): knot 7 & 0.94 & 0.01 & 0.00 \\ 
Sex : M * Spline(age): knot 8 & 0.84 & 0.02 & 0.00 \\ 
\bottomrule
\end{longtable}

\newpage

\hypertarget{references}{%
\section*{References}\label{references}}
\addcontentsline{toc}{section}{References}

\hypertarget{refs}{}
\begin{CSLReferences}{1}{0}
\leavevmode\vadjust pre{\hypertarget{ref-abe_japans_2013}{}}%
Abe, Shinzo. 2013. {``Japan's Strategy for Global Health Diplomacy: Why
It Matters.''} \emph{The Lancet} 382 (9896): 915--16.
\url{https://doi.org/10.1016/S0140-6736(13)61639-6}.

\leavevmode\vadjust pre{\hypertarget{ref-bogaert_use_2018}{}}%
Bogaert, Petronille, Herman Van Oyen, Isabelle Beluche, Emmanuelle
Cambois, and Jean-Marie Robine. 2018. {``The Use of the Global Activity
Limitation {Indicator} and Healthy Life Years by Member States and the
{European} {Commission}.''} \emph{Archives of Public Health} 76 (1): 30.
\url{https://doi.org/10.1186/s13690-018-0279-z}.

\leavevmode\vadjust pre{\hypertarget{ref-euro-reves_selection_2000}{}}%
Euro-REVES, Carol Jagger, Viviana Egidi, and Jean Marie Robine. 2000.
{``Selection of a {Coherent} {Set} of {Health} {Indicators}.''} Final
draft. Euro-REVES.
\url{https://ec.europa.eu/health/ph_projects/1998/monitoring/fp_monitoring_1998_frep_03_en.pdf}.

\leavevmode\vadjust pre{\hypertarget{ref-eurostat_healthy_2020}{}}%
Eurostat. 2020. {``Healthy Life Years by Sex (from 2004 Onwards)
(Hlth\_hlye).''} \emph{Eurostat Database}.
\url{https://ec.europa.eu/eurostat/databrowser/view/HLTH_HLYE/default/table?lang=en}.

\leavevmode\vadjust pre{\hypertarget{ref-fries_aging_1980}{}}%
Fries, James F. 1980. {``Aging, {Natural} {Death}, and the {Compression}
of {Morbidity}.''} \emph{New England Journal of Medicine} 303 (3):
130--35. \url{https://doi.org/10.1056/NEJM198007173030304}.

\leavevmode\vadjust pre{\hypertarget{ref-gruenberg_failures_2005}{}}%
Gruenberg, Ernest M. 2005. {``The {Failures} of {Success}.''}
\emph{Milbank Quarterly} 83 (4): 779--800.
\url{https://doi.org/10.1111/j.1468-0009.2005.00400.x}.

\leavevmode\vadjust pre{\hypertarget{ref-guibert_mesure_2018}{}}%
Guibert, Quentin, Frédéric Planchet, and Michaël Schwarzinger. 2018a.
{``Mesure de l'espérance de Vie En Dépendance Totale En France.''}
\emph{Bulletin Français d'Actuariat} 18 (35): 133--59.
\url{https://hal.archives-ouvertes.fr/hal-02055147}.

\leavevmode\vadjust pre{\hypertarget{ref-guibert_mesure_2018-1}{}}%
---------. 2018b. {``Mesure de l'espérance de Vie Sans Dépendance Totale
En {France} Métropolitaine.''} \emph{Bulletin Français d'Actuariat} 18
(35): 85--109. \url{https://hal.archives-ouvertes.fr/hal-02055147}.

\leavevmode\vadjust pre{\hypertarget{ref-head_socioeconomic_2019}{}}%
Head, Jenny, Holendro Singh Chungkham, Martin Hyde, Paola Zaninotto,
Kristina Alexanderson, Sari Stenholm, Paula Salo, et al. 2019.
{``Socioeconomic Differences in Healthy and Disease-Free Life Expectancy
Between Ages 50 and 75: A Multi-Cohort Study.''} \emph{European Journal
of Public Health} 29 (2): 267--72.
\url{https://doi.org/10.1093/eurpub/cky215}.

\leavevmode\vadjust pre{\hypertarget{ref-insee_situation_2022}{}}%
INSEE. 2022. {``La Situation Démographique En 2020.''} \emph{INSEE
Reports}.
\url{https://www.insee.fr/fr/statistiques/6327226?sommaire=6327254}.

\leavevmode\vadjust pre{\hypertarget{ref-insee_esperances_2023}{}}%
---------. 2023. {``Espérances de Vie à Différents Âges.''} \emph{INSEE
Reports}.
\url{https://www.insee.fr/fr/outil-interactif/6794598/EVDA/DEPARTMENTS}.

\leavevmode\vadjust pre{\hypertarget{ref-jagger_international_2020}{}}%
Jagger, Carol, Eileen M. Crimmins, Yasuhiko Saito, Renata Tiene De
Carvalho Yokota, Herman Van Oyen, and Jean-Marie Robine, eds. 2020.
\emph{International {Handbook} of {Health} {Expectancies}}. Vol. 9.
International {Handbooks} of {Population}. Cham: Springer International
Publishing. \url{https://doi.org/10.1007/978-3-030-37668-0}.

\leavevmode\vadjust pre{\hypertarget{ref-kempen_assessment_1996}{}}%
Kempen, Gertrudis I. J. M., Nardi Steverink, Johan Ormel, and Dorly J.
H. Deeg. 1996. {``The {Assessment} of {ADL} Among {Frail} {Elderly} in
an {Interview} {Survey}: {Self}-Report Versus {Performance}-{Based}
{Tests} and {Determinants} of {Discrepancies}.''} \emph{The Journals of
Gerontology Series B: Psychological Sciences and Social Sciences} 51B
(5): P254--60. \url{https://doi.org/10.1093/geronb/51B.5.P254}.

\leavevmode\vadjust pre{\hypertarget{ref-kim_review_2022}{}}%
Kim, Young-Eun, Yoon-Sun Jung, Minsu Ock, and Seok-Jun Yoon. 2022. {``A
{Review} of the {Types} and {Characteristics} of {Healthy} {Life}
{Expectancy} and {Methodological} {Issues}.''} \emph{Journal of
Preventive Medicine and Public Health} 55 (1): 1--9.
\url{https://doi.org/10.3961/jpmph.21.580}.

\leavevmode\vadjust pre{\hypertarget{ref-klein_handbook_2016}{}}%
Klein, John P., Hans C. Van Houwelingen, Joseph G. Ibrahim, and Thomas
H. Scheike, eds. 2016. \emph{Handbook of {Survival} {Analysis}}. Boca
Raton, FL: Chapman; Hall/CRC. \url{https://doi.org/10.1201/b16248}.

\leavevmode\vadjust pre{\hypertarget{ref-krause_what_1994}{}}%
Krause, Neal M., and Gina M. Jay. 1994. {``What {Do} {Global}
{Self}-{Rated} {Health} {Items} {Measure}?:''} \emph{Medical Care} 32
(9): 930--42. \url{https://doi.org/10.1097/00005650-199409000-00004}.

\leavevmode\vadjust pre{\hypertarget{ref-lagstrom_diet_2020}{}}%
Lagström, Hanna, Sari Stenholm, Tasnime Akbaraly, Jaana Pentti, Jussi
Vahtera, Mika Kivimäki, and Jenny Head. 2020. {``Diet Quality as a
Predictor of Cardiometabolic Disease--Free Life Expectancy: The
{Whitehall} {II} Cohort Study.''} \emph{The American Journal of Clinical
Nutrition} 111 (4): 787--94. \url{https://doi.org/10.1093/ajcn/nqz329}.

\leavevmode\vadjust pre{\hypertarget{ref-martinussen_dynamic_2006}{}}%
Martinussen, Torben, and Thomas H. Scheike. 2006. \emph{Dynamic
Regression Models for Survival Data}. Statistics for Biology and Health.
New York, N.Y: Springer.

\leavevmode\vadjust pre{\hypertarget{ref-peersman_gender_2012}{}}%
Peersman, Wim, Dirk Cambier, Jan De Maeseneer, and Sara Willems. 2012.
{``Gender, Educational and Age Differences in Meanings That Underlie
Global Self-Rated Health.''} \emph{International Journal of Public
Health} 57 (3): 513--23.
\url{https://doi.org/10.1007/s00038-011-0320-2}.

\leavevmode\vadjust pre{\hypertarget{ref-r_core_team_r_2022}{}}%
R Core Team. 2022. \emph{R: {A} {Language} and {Environment} for
{Statistical} {Computing}}. Vienna, Austria: R Foundation for
Statistical Computing. \url{https://www.R-project.org/}.

\leavevmode\vadjust pre{\hypertarget{ref-robine_creating_2003}{}}%
Robine, Jean-Marie. 2003. {``Creating a Coherent Set of Indicators to
Monitor Health Across {Europe}: {The} {Euro}-{REVES} 2 Project.''}
\emph{The European Journal of Public Health} 13 (Supplement 1): 6--14.
\url{https://doi.org/10.1093/eurpub/13.suppl_1.6}.

\leavevmode\vadjust pre{\hypertarget{ref-sanders_measuring_1964}{}}%
Sanders, Barkev S. 1964. {``Measuring Community Health Levels.''}
\emph{American Journal of Public Health and the Nation's Health} 54 (7):
1063--70. \url{https://doi.org/10.2105/AJPH.54.7.1063}.

\leavevmode\vadjust pre{\hypertarget{ref-schwarzinger_etude_2018}{}}%
Schwarzinger, Michaël. 2018. {``Etude {QalyDays} : Données Source Et
Retraitements Pour l'étude Du Risque Perte d'autonomie.''}
\emph{Bulletin Français d'Actuariat} 18 (35).
\url{http://www.ressources-actuarielles.net/EXT/IA/sitebfa.nsf/5aa33af8a2183d96c125787a006b34c6/9d16f73ea73a00e6c12583a800239fdf?OpenDocument}.

\leavevmode\vadjust pre{\hypertarget{ref-schwarzinger_contribution_2018}{}}%
Schwarzinger, Michaël, Bruce G Pollock, Omer S M Hasan, Carole Dufouil,
Jürgen Rehm, S Baillot, Q Guibert, F Planchet, and S Luchini. 2018.
{``Contribution of Alcohol Use Disorders to the Burden of Dementia in
{France} 2008--13: A Nationwide Retrospective Cohort Study.''} \emph{The
Lancet Public Health} 3 (3): e124--32.
\url{https://doi.org/10.1016/S2468-2667(18)30022-7}.

\leavevmode\vadjust pre{\hypertarget{ref-stenholm_body_2017}{}}%
Stenholm, Sari, Jenny Head, Ville Aalto, Mikai Kivimäki, Ichiro Kawachi,
Marie Zins Zins, Marcel Goldberg, et al. 2017. {``Body Mass Index as a
Predictor of Healthy and Disease-Free Life Expectancy Between Ages 50
and 75: A Multicohort Study.''} \emph{International Journal of Obesity}
41 (5): 769--75. \url{https://doi.org/10.1038/ijo.2017.29}.

\leavevmode\vadjust pre{\hypertarget{ref-therneau_package_2023}{}}%
Therneau, Terry M. 2023. \emph{A {Package} for {Survival} {Analysis} in
{R}}. \url{https://CRAN.R-project.org/package=survival}.

\leavevmode\vadjust pre{\hypertarget{ref-world_health_organization_international_2015}{}}%
World Health Organization. 2015. \emph{International Statistical
Classification of Diseases and Related Health Problems, 10th Revision}.
Fifth Edition. Geneva: World Health Organization.
\url{https://apps.who.int/iris/handle/10665/246208}.

\leavevmode\vadjust pre{\hypertarget{ref-world_health_organization_world_2023}{}}%
---------. 2023. {``World Health Statistics: {Monitoring} Health for the
{SGDs}, Sustainable Development Goals.''} Geneva: World Health
Organization.

\end{CSLReferences}

\bibliographystyle{unsrt}
\bibliography{MyLibrary.bib}

\end{document}